\documentclass[twocolumn,showpacs,preprintnumbers,amsmath,amssymb,floatfix]{revtex4}

\usepackage{graphicx}
\usepackage{dcolumn}
\usepackage{bm}

\begin{document}

\title{On the Formation and Progenitor of PSR\,J0737-3039: New
  Constraints on the Supernova Explosion Forming Pulsar~B}   

\author{B. Willems}\email{b-willems@northwestern.edu}
\author{J. Kaplan}\email{j-kaplan-1@northwestern.edu}
\author{T. Fragos}\email{tassosfragos@northwestern.edu}
\author{V. Kalogera}\email{vicky@northwestern.edu}
\affiliation{Northwestern University, Department of Physics and
  Astronomy, 2145 Sheridan Road, Evanston, IL 60208, USA}
\author{K. Belczynski}\email{kbelczyn@nmsu.edu}
\affiliation{New Mexico State University, Department of Astronomy,
  1320 Frenger Mall, Las Cruces, NM 88003, USA \\ 
  Tombaugh Fellow}
\begin{abstract}
We investigate the formation of the double pulsar PSR\,J0737-3039 and examine
its most likely progenitors, taking into account the most recent and all
currently available observational constraints. We show that the most
likely kick velocity and progenitor parameters depend strongly on the
consideration of the full five-dimensional probability distribution
function for the magnitude and direction of the kick velocity imparted
to pulsar~B at birth, the mass of pulsar~B's pre-supernova helium star
progenitor, and the pre-supernova orbital separation rather than marginalized one- or two-dimensional distributions for the kick velocity and progenitor mass. The priors that enter the analysis are the age
of the system, the minimum helium star mass required to form a neutron
star, the transverse systemic velocity, and the treatment of the
unknown radial velocity. Since the latter cannot be measured observationally,
we adopt a statistical approach and use theoretical
radial-velocity distributions obtained from population synthesis
calculations for coalescing double neutron stars. We find that when the minimum
pre-supernova helium star mass required for neutron star formation is
assumed to be $2.1\,M_\odot$, the most likely kick velocity ranges from
$70\,{\rm km\,s^{-1}}$ to $180\,{\rm km\,s^{-1}}$. When,
on the other hand, masses lower than $2.1\,M_\odot$ are allowed as neutron star
progenitors, the most
likely kick velocity can reach very low values (as low as a few ${\rm km\,s^{-1}}$),
although the majority of the models still yield most likely
kick velocities of $50\,{\rm km\,s^{-1}}$ to $170\,{\rm km\,s^{-1}}$.
Hence, we agree with \citet{2005PhRvL..94e1102P} that the observed system
properties, including the low transverse systemic velocity, {\em can} indeed be
compatible with low progenitor masses and low kick velocities. Equally important though, we show that this is not 
the only likely formation path of pulsar~B,  due to
the role of different prior assumptions that are necessary in the analysis. Moreover, in contrast to earlier claims in the  literature, we show that the
proximity of the double pulsar to the Galactic plane and the small proper
motion do not pose stringent constraints on the kick velocity and
progenitor mass of pulsar~B at all. Instead, the constraints imposed by
the current orbital semi-major axis and eccentricity and the orbital
dynamics of asymmetric supernova explosions turn out to be much more
restrictive. We conclude that without further
knowledge of the priors, the currently available observational constraints
cannot be used to unambiguously favor a specific core-collapse and neutron
star formation mechanism. Both electron capture and neutrino-driven
supernovae therefore remain viable formation mechanisms for pulsar~B.
\end{abstract}


\maketitle

\section{Introduction}

Since its discovery in 2003 \citep{2003Natur.426..531B,
2004Sci...303.1153L}, the double pulsar PSR\,J0737-3039 has proven to
be an extraordinary system for the study of both fundamental physics
and general relativity \citep{2005MNRAS.361.1243P,
2005astro.ph..3386K}. The system is highly relativistic with an
orbital period of 2.4\,hrs and an eccentricity of 0.0878, making it
the tightest known double neutron star (DNS) binary known to date. The
system also harbors the fastest known DNS pulsar (22.7\,ms), has the
largest apsidal-motion rate of all presently known DNSs
($16.9^\circ\,{\rm yr^{-1}}$), and shows eclipses of the fast pulsar
by the magnetosphere of the slow one (orbital inclination $i =
87^\circ \pm 3^\circ$) \citep{2003Natur.426..531B,
2004Sci...303.1153L}. Its short merger time of $\sim 85$\,Myr has
furthermore led to a significant upward revision for the merger rate
of coalescing DNSs \citep{2004ApJ...601L.179K, 2004ApJ...614L.137K}.

A central topic in understanding the physics of the double pulsar is
the study of its origin and evolution.  According to our current
knowledge (for more details see \citep{1991PhR...203....1B,
1995ApJ...440..270B, 2002ApJ...572..407B, 2003astro.ph..3456T}), DNSs
form from primordial binaries in which, possibly after some initial
mass-transfer phase, both component stars have masses in excess of
$\sim 8-12\,M_\odot$. After the primary explodes in a supernova (SN)
explosion to form the first neutron star (NS), the binary enters a
high-mass X-ray binary phase in which the NS accretes matter from the
wind of its companion. The phase ends when the companion star evolves
of the main sequence and engulfs the NS in its expanding envelope. The
NS then spirals in towards the core of the companion until enough
orbital energy is transferred to the envelope to expel it from the
system. When the envelope is ejected, the binary consists of the NS
and the stripped-down helium core of its former giant companion,
orbiting each other in a tight orbit. If the NS is able to accrete
during its rapid inspiral, a first "recycling" may take place during
which it is spun up to millisecond periods. The helium star then
evolves further until it, in turn, explodes and forms a NS. Depending
on the mass of the helium star and the size of the orbit at the time
of the explosion, the SN may be preceded by a second recycling phase
when the helium star fills its Roche lobe and transfers mass to the
NS \citep{2001ApJ...550L.183B, 2002ApJ...571L.147B, 2002ApJ...572..407B}.

Shortly after the discovery of PSR\,J0737-3039,
\citet{2004MNRAS.349..169D} and \citet{2004ApJ...603L.101W} (hereafter
Paper~I) independently derived that, right before the second SN
explosion, the binary was so tight that the helium star must have been
overflowing its critical Roche lobe. This conclusion, for the first
time, strongly confirmed the above formation channel for DNS
binaries. The observed 22.7\,ms pulsar (hereafter PSR\,J0737-3039A or
pulsar~A) then corresponds to the first-born NS, and its 2.8\,s pulsar
companion (hereafter PSR\,J0737-3039B or pulsar~B) to the second-born
NS.

In a follow-up study to Paper~I, \citet{2004ApJ...616..414W}
(hereafter Paper~II) extended the stellar and binary evolution
constraints on the formation of pulsar~B to include constraints based
on the system's position and motion in the Galaxy. While the addition
of proper motion measurements to the list of observational
constraints, at first sight, added a significant piece of information
on the formation and evolution of the double pulsar system, the
unknown direction of the proper motion and the unknown radial velocity
significantly hindered the derivation of more stricter progenitor
constraints than obtained in Paper~I.

Since Paper~II, the available observational constraints have undergone
significant revisions. Most notably, proper motion estimates, which
form a vital part in constraining the kick imparted to pulsar~B at
birth, have decreased from $\sim 141\,{\rm km\,s^{-1}}$ down to
$\sim 30\,{\rm km\,s^{-1}}$ or less \citep{2004ApJ...609L..71R,
2005ApJ...623..392C, 2005astro.ph..3386K}. In addition, rough limits
on the spin-orbit misalignment of pulsar~A have recently been derived
by \citet{2005ApJ...621L..49M}, which again add to the constrains on
pulsar~B's natal kick velocity.  From a theoretical point of view, it
has furthermore been argued that pulsar~B may have had an unusually
low-mass progenitor ($\sim 1.45\,M_\odot$) which formed a NS through a
new type of gravitational collapse accompanied by an unusually low
kick velocity ($\lesssim 30\,{\rm km\,s^{-1}}$)
\citep{2005PhRvL..94e1102P, 2005astro.ph.10584P}.

Our aim in this paper is to investigate the system's recent
evolutionary history and examine the necessity of this new type of
gravitational collapse to explain the formation of pulsar~B.  We
incorporate the most up-to-date observational information on
PSR\,J0737-3039, including the $30\,{\rm km\,s^{-1}}$ upper limit on
the proper motion derived by \citet{2005astro.ph..3386K}. 
A significant new element compared to Papers~I and~II, which  
focused on the extent of the available parameter space, is the
examination of the full multi-dimensional probability distribution
functions (PDF) for the magnitude and direction of pulsar~B's natal
kick velocity, the mass of pulsar~B's pre-SN helium star progenitor,
and the orbital separation right before the formation of
pulsar~B. Other significant improvements are the incorporation of
theoretical distribution functions for the unknown radial velocity by
means of binary population synthesis calculations for coalescing DNSs,
the investigation of projection and marginalization effects of the
multi-dimensional PDF, and the examination of the role of the prior
assumptions.

The plan of the paper is as follows. In \S\,II, we briefly summarize
the observationally inferred properties of the double pulsar relevant
to this investigation. In \S\,III, we present the basic assumptions
and outline the method used to derive the formation history of
pulsar~B. Details of the various steps adopted in the investigation
and their overall role in the analysis are presented in \S\,IV and
\S\,V. In \S\,VI, we present the most likely progenitor and formation
parameters for PSR\,J0737-3039B and systematically explore the
sensitivity of the parameters to the adopted assumptions. The final
two sections, \S\,VII and \S\,VIII, are devoted to a comparison with
previous investigations and summarizing remarks.

\section{Observational constraints}

The main properties of the double pulsar system and its component stars
relevant to this investigation are summarized in Table~\ref{prop}. The
principal differences with the observational constraints used in Paper~II
are the considerably lower estimate for the systemic velocity
perpendicular to the line-of-sight, and the availability of limits on
the spin-orbit misalignment angle of pulsar~A.

\begin{table*}
\caption{Physical properties of PSR\,J0737-3039. \label{prop}}
\begin{ruledtabular}
\begin{tabular}{lccc}
\textbf{Parameter}                        & \textbf{Notation} & \textbf{Value} & \textbf{Reference} \\
\hline 
Distance (pc)                             & $d$               & 600    & \citep{2003Natur.426..531B, 2004Sci...303.1153L} \\
Galactic longitude (J2000) (deg)          & $l$               & 245.2  & \citep{2003Natur.426..531B, 2004Sci...303.1153L} \\
Galactic latitude (J2000) (deg)           & $b$               & -4.5   & \citep{2003Natur.426..531B, 2004Sci...303.1153L} \\
Proper motion (km\,s$^{-1}$)              & $V_t$             & $< 30$ & \citep{2005astro.ph..3386K} \\
Spin period of pulsar A (ms)              & $P_A$             & 22.7   & \citep{2003Natur.426..531B, 2004Sci...303.1153L} \\
Spin period of pulsar B (s)               & $P_B$             &  2.8   & \citep{2003Natur.426..531B, 2004Sci...303.1153L} \\
Mass of pulsar A ($M_\odot$)              & $M_A$             & 1.34   & \citep{2003Natur.426..531B, 2004Sci...303.1153L} \\
Mass of pulsar B ($M_\odot$)              & $M_B$             & 1.25   & \citep{2003Natur.426..531B, 2004Sci...303.1153L} \\
Characteristic age of pulsar A (Myr)      & $\tau_A$          & 210    & \citep{2003Natur.426..531B, 2004Sci...303.1153L} \\
Characteristic age of pulsar B (Myr)      & $\tau_B$          & 50     & \citep{2003Natur.426..531B, 2004Sci...303.1153L} \\
Spin-down age of pulsar A (Myr)           & $\tau_{\rm sd}$   & 100    & \citep{2003Natur.426..531B, 2004Sci...303.1153L} \\
Spin-orbit misalignment of pulsar~A (deg) & $\lambda$         & $< 60$ or $> 120$ & \citep{2005ApJ...621L..49M} \\
Orbital period (hours)                    & $P_{\rm orb}$     & 2.4    & \citep{2003Natur.426..531B, 2004Sci...303.1153L} \\
Orbital semi-major axis ($R_\odot$)       & $A_c$             & 1.26   & \citep{2003Natur.426..531B, 2004Sci...303.1153L} \\
Orbital eccentricity                      & $e_c$             & 0.0878 & \citep{2003Natur.426..531B, 2004Sci...303.1153L} \\
\end{tabular}
\end{ruledtabular}
\end{table*}

The kinematic properties of PSR\,J0737-3039 have undergone significant
revision since \citet{2004ApJ...609L..71R} used interstellar
scintillation measurements and inferred a velocity component
perpendicular to the line-of-sight of $\sim 141\,{\rm
km\,s^{-1}}$. \citet{2005ApJ...623..392C} derived a reduced velocity
of $\sim 66\,{\rm km\,s^{-1}}$ by incorporating anisotropies of the
interstellar medium in the interstellar scintillation model. This
reduced velocity, however, strongly depends on the adopted anisotropy
model. \citet{2005astro.ph..3386K}, on the other hand, used pulsar
timing measurements to derive a firm model-independent upper limit of
$30\,{\rm km\,s^{-1}}$ on the transverse velocity. In view of these
revisions, we update the kinematic properties of PSR\,J0737-3039
derived in Paper~II, adopting the most recent constraint of $30\,{\rm
km\,s^{-1}}$ as an upper limit to the velocity component transverse to
the line-of-sight. 

The kick imparted to pulsar~B at birth is expected to tilt the orbital
plane and misalign pulsar~A's spin axis with respect to the post-SN
orbital angular momentum axis \citep[e.g.][]{2000ApJ...541..319K}. The
degree of misalignment depends on both the magnitude and the direction
of the kick, and therefore yields a valuable piece of information on
pulsar~B's natal kick velocity (see \S\,\ref{asymSN} for more
details). In Paper~II, we showed that for the pre- and post-SN orbital
parameters compatible with all available observational constraints,
misalignment angles between approximately $70^\circ$ and $110^\circ$
are highly unlikely. \citet{2005ApJ...621L..49M} derived observational
constraints on the spin-orbit misalignment based on the stability of
pulsar~A's mean pulse profile over a time span of 3 years. The authors
concluded the angle to be smaller than $\sim 60^\circ$ or larger than
$\sim 120^\circ$, in agreement with the theoretical predictions of
Paper~II. In the present paper, we therefore include the limits on
pulsar~A's spin-orbit misalignment inferred by
\citet{2005ApJ...621L..49M} in the list of available observational
constraints.

Among the constraints inferred from observations, the most uncertain
parameter affecting the reconstruction of the system's formation and
evolutionary history is its age. Characteristic ages, defined
as half the ratio of the spin period to the spin-down rate, are the
most commonly adopted age estimators, but are known to be quite
unreliable \citep[e.g.][]{2003ApJ...593L..31K}. In the case of
PSR\,J0737-3039, the spin evolution of pulsar~B is furthermore very
likely affected by torques exerted by pulsar~A on pulsar~B \citep{2004MNRAS.353.1095L}, adding to
the uncertainties of its age. \citet{2004astro.ph..4274L} therefore
derived an alternative age estimate by noting that, according to the
standard DNS formation channel, the time expired since the end of
pulsar~A's spin-up phase should be equal to the time expired since the
birth of pulsar~B. By combining this property with a selection of
different spin-down models, the authors derived a {\em most likely}
age of 30-70\,Myr, but were unable to firmly exclude younger and older
ages. A third age estimate can be obtained by assuming that pulsar~A
was recycled to the maximum spin rate and calculating the time
required for it to spin down to the currently observed value. This
so-called spin-down age provides an upper limit to the age of the
system of 100\,Myr \citep{2003Natur.426..531B}. In view of these
uncertainties, we derive constraints on the formation and evolution of
PSR\,J0737-3039 for three different sets of age assumptions: (i) $0
\le \tau \le 100$\,Myr (the age range adopted in Papers~I and~II),
(ii) $30 \le \tau \le 70$\,Myr (the most likely age range derived by
\citet{2004astro.ph..4274L}), and (iii) $\tau \simeq 50$\,Myr (the
characteristic age of pulsar~B).

\section{Basic assumptions and outline of the calculation}
\label{basic}

Our goal in this investigation is to reconstruct the evolutionary
history of PSR\,J0737-3039 and constrain its properties at the
formation time of pulsar~B. More specifically, our aim is to derive a
PDF for the magnitude and direction of the kick velocity imparted to
pulsar~B at birth, the mass of pulsar~B's progenitor
immediately before it explodes in a SN explosion, and the orbital
separation of the component stars right before the SN explosion.
Adopting the standard DNS formation channel, the progenitor of the
double pulsar right before the formation of pulsar~B, consists of the
first-born NS, pulsar~A, and the helium star progenitor of the
second-born NS. Since the formation of the second NS is preceded by
one or more mass-transfer phases, tidal forces can safely be assumed
to circularize the orbit prior to the formation of pulsar~B.  For the
remainder of the paper, we refer to the times right before and right
after the formation of pulsar~B by pre-SN and post-SN,
respectively. The subsequent analysis consists of four major parts.

Firstly, the motion of the system in the Galaxy is traced back in time
up to a maximum age of $100$\,Myr. The goal of this calculation is to
derive the position and center-of-mass velocity of the binary right
after the formation of pulsar~B, and use this information to constrain
the kick imparted to it at birth. In order to determine possible birth
sites, our analysis is supplemented with the assumptions that the
primordial DNS progenitor was born close to the Galactic plane, and
that the first SN explosion did not kick the binary too far out of it.
The first assumption is conform with our current knowledge that
massive stars form and live close to the Galactic plane (their typical
scale height is $\simeq 50$--70\,pc). The second assumption
on the other hand neglects a small fraction of systems formed with
high space velocities \citep{2002ApJ...574..364P, 2002ApJ...572..407B}
\footnote{From Fig.~6 in \citet{2002ApJ...572..407B} it follows that
less than 2\,\% of the systems have a peculiar velocity larger than
100\,km\,s$^{-1}$ after the first SN explosion.}. Under these
assumptions, the binary is still close to the Galactic plane at the
formation time of pulsar~B. Possible birth sites can thus be obtained
by calculating the motion in the Galaxy backwards in time and looking
for crossings of the orbit with the Galactic mid-plane. The times in
the past at which the plane crossings occur yield kinematic estimates
for the age of the DNS, while comparison of the system's total
center-of-mass velocity with the local Galactic rotational velocity at
the birth sites yields the system's post-SN {\em peculiar} velocity.

Secondly, the orbital semi-major axis and eccentricity right after the
SN explosion forming pulsar~B are determined by integrating the
equations governing the evolution of the orbit due to gravitational
wave emission backwards in time. The integration is performed for each
Galactic plane crossing found from the Galactic motion
calculations. Since each crossing yields a different kinematic age,
and thus a different endpoint of the reverse orbital evolution calculation,
the post-SN parameters are a function of the considered Galactic plane
crossing.

Thirdly, the conservation laws of orbital energy and angular momentum
are used to map the post-SN binary parameters to the pre-SN ones. The
mapping depends on the kick velocity imparted to pulsar~B at birth and
results in constraints on the magnitude and direction of the kick
velocity, the mass of pulsar~B's pre-SN helium star progenitor, and
the pre-SN orbital separation. The pre-SN orbital eccentricity is
assumed to be zero, as expected from strong tidal forces operating on
the binary during the mass-transfer phase(s) responsible for spinning
up pulsar~A. The kick and pre- and post-SN binary parameters are then
furthermore subjected to the requirements that they be consistent with
the post-SN peculiar velocity obtained from the Galactic motion
calculations and with the observationally inferred spin-orbit
misalignment angle of pulsar~A. The latter constraint requires the
additional assumption that tidal forces align the pre-SN rotational
angular momentum vector of pulsar~A with the pre-SN orbital angular
momentum vector.

Fourthly, the solutions of the conservation laws of orbital energy and
angular momentum are weighted according to the likelihood that they
lead to the system's currently observed position and velocity in the
Galaxy. A PDF of the admissible kick velocity and progenitor
parameters is then constructed by binning the solutions in a
multi-dimensional ``rectangular'' grid. The maximum of the resulting
PDF yields the most likely magnitude and direction of the kick
velocity imparted to pulsar~B at birth, mass of pulsar~B's pre-SN
helium star progenitor, and pre-SN orbital separation. We also
investigate the sensitivity of the PDF to the adopted assumptions by
systematically varying the underlying assumptions, such as, e.g., the
age and the magnitude of the transverse systemic velocity component.

\section{Kinematic age and history}

\subsection{Galactic motion}
\label{motion}

Tracing the motion of PSR\,J0737-3039 back in time requires the
knowledge of both its present-day position and 3-dimensional space
velocity. Unfortunately, no method is presently available to measure
radial velocities of DNSs, so that the knowledge of their space
velocity is limited to the component perpendicular to the
line-of-sight (transverse velocity). At present, only an upper
limit is available for the double pulsar: $V_t \lesssim 30\,{\rm
km\,s^{-1}}$. The direction of the motion in the plane perpendicular
to the line-of-sight is still unknown. Therefore, as in
Paper~II, we explore the system's kinematic history in terms of two
unknown parameters: (i) the radial component $V_r$ of the present-day
systemic velocity, and (ii) the orientation $\Omega$ of the transverse
velocity in the plane perpendicular to the line-of-sight. Since the
results presented in this paper do not depend on the exact definition
of $\Omega$, we refer the reader to Paper~II for a more detailed
discussion and definition. 

The motion of the system in the Galaxy is calculated with respect to a
right-handed frame of reference $OXYZ$ with origin at the Galactic
center and with the $XY$-plane coinciding with the Galactic
mid-plane. The $X$-axis is chosen such that the Sun is located in the
$XZ$-plane, and the positive direction of the $Y$-axis such that it
coincides with the direction of the Galactic rotational velocity at
the position of the Sun. The velocity components $V_X$, $V_Y$, $V_Z$
with respect to this frame of reference are obtained from the radial
and transverse velocity components $V_r$ and $V_t$ by a standard
transformation of vector components, similar to the one adopted in
\S\,2.2 of Paper~II. We note that because PSR\,J0737-3039 is located
only 20\,pc below the Galactic mid-plane \footnote{This is assuming the
Sun is located at a height of 30\,pc above the Galactic plane
\citep[see, e.g.,][]{2001ApJ...553..184C}}, the unknown radial
velocity $V_r$ mainly affects the determination of $V_X$ and $V_Y$.

As in Paper~II, we calculate the motion of the system in the Galaxy
backwards in time using the Galactic potential of
\citet{1987AJ.....94..666C} with updated model parameters derived by
\citet{1989MNRAS.239..571K} \footnote{In Paper~II, the motion of the
system in the Galaxy backwards in time was accidentally calculated for
the wrong set of current Galactic coordinates. Due to the system's
proximity to the Sun ($d \sim 0.6\,{\rm kpc}$) this error mainly
affected the appearance of Figs. 3--5, but not the derived ranges of
kinematic ages and post-SN peculiar velocities. Correspondingly, the
derived constraints on the mass of pulsar~B's pre-SN helium star
progenitor, the pre-SN orbital separation, the magnitude and direction
of pulsar~B's natal kick velocity, and the probability distributions
for the kick magnitude and spin-orbit misalignment angle are also
largely unaffected.}. Since the present-day kinematical constraints
only provide an upper limit of 30\,km\,$s^{-1}$ on the transverse
systemic velocity and the calculation of the past orbit requires
precise starting values for both the present-day position and the
present-day velocity, we calculate the motion backwards in time for
two specific velocities: $V_t=10$\,km\,s$^{-1}$ and
$V_t=30$\,km\,s$^{-1}$. For each of these $V_t$, we integrate the
equations of motion up to 100\,Myr back in time for all possible
$V_r$-values between -1500\,km\,s$^{-1}$ and 1500\,km\,s$^{-1}$ (in
steps of 10\,km\,s$^{-1}$), and $\Omega$-values between $0^\circ$ and
$360^\circ$ (in steps of $5^\circ$). Although extremely high radial
velocities in excess of $\sim 1000$\,km\,s$^{-1}$ may seem rather
unlikely, they cannot be firmly excluded by the presently known
observational constraints (see Paper~II). The possibility of a high
space velocity is also supported by observations of high-velocity
single radio pulsars such as PSR\,B2224+65 (with $V_t \gtrsim 800\,
{\rm km\,s^{-1}}$ \citep{1993Natur.362..133C}) and PSR\,B1508+55 (with
$V_t = 1083^{+103}_{-90}\, {\rm km\,s^{-1}}$
\citep{2005ApJ...630L..61C}). Moreover, as many of these high-velocity
pulsars are expected to quickly escape the Galaxy, there is a strong
bias against finding them. The fraction of radio pulsars formed with
high space velocities is therefore not well constrained in the current
pulsar sample.

In order to account for the possibility that PSR\,J0737-3039 has a
large radial velocity and examine the effect of incorporating large
radial velocities in the analysis, we weigh each considered radial
velocity according to a pre-determined radial velocity
distribution. For this purpose, we performed a population synthesis
study of Galactic DNSs, including their kinematic evolution in the
potential of \citet{1987AJ.....94..666C}. Theoretical radial velocity
distributions are then obtained by taking a snapshot of the population
at the current epoch and determining the radial velocity for each DNS
in the sample. The resulting PDFs are found to be represented well by
Gaussian distributions with means of 0\,km\,s$^{-1}$ and velocity
dispersions of 60--200\,km\,s$^{-1}$, depending on the adopted
population synthesis assumptions (see \S\,\ref{pvrad} for more
details). For comparison, we also consider a uniform radial velocity
distribution in which all radial velocities between
-1500\,km\,s$^{-1}$ and 1500\,km\,s$^{-1}$ are equally
probable \footnote{This uniform distribution is what we also assumed in
Paper~II. \citet{2005astro.ph.10584P} incorrectly state that the
results presented in Paper~I are dictated by the implicit assumption
that the system is moving with a large radial velocity towards us. The
kick velocity and progenitor parameters derived in Paper~I are,
however, based solely on stellar and binary evolution
considerations. Constraints on the proper motion did not become
available until the writing of Paper~II in which we presented the
possible kick velocity and progenitor parameters {\em as a function}
of the unknown radial velocity. In that paper, we considered the full
range of radial velocities, from small to large, as having a flat
distribution function.}. It is clear that, contrary to what is stated by \citet{2006astro.ph..3649P}, we do not assume the system to be moving almost exactly towards us with a very large radial velocity $V_r$. For the other unknown parameter, the orientation angle $\Omega$ of the transverse velocity component in the plane perpendicular to the line-of-sight, we assume a uniform
distribution between $0^\circ$ and $360^\circ$.

As noted in the previous section, it is reasonable to assume that
PSR\,J0737-3039 was born close to the Galactic plane, so that the
intersections of the past orbits with the $Z=0$ plane yield possible
birth sites as functions of $V_r$ and $\Omega$. The number of Galactic
plane crossings associated with each $V_r$ and $\Omega$, within the
adopted age limit of 100\,Myr, can be anywhere between 1 and 5. The
times in the past at which the system crosses the Galactic plane
furthermore yield kinematic estimates for the age of the DNS, while
subtraction of the Galactic rotational velocity from the total
systemic velocity at the birth sites yields the peculiar velocity
right after the formation of pulsar~B.

\begin{figure*}
\resizebox{14.5cm}{!}{\includegraphics{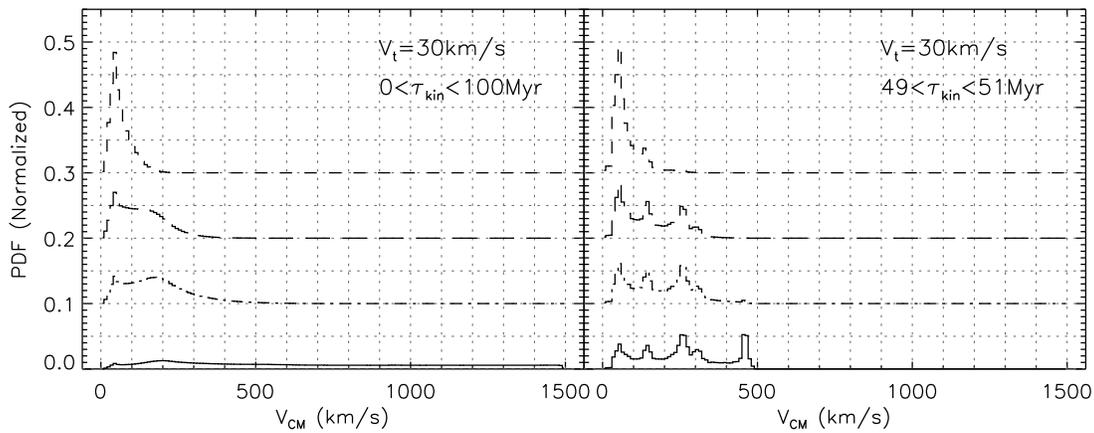}}
\caption{Distribution of post-SN peculiar velocities for a present-day
  transverse velocity of 30\,km\,s$^{-1}$, kinematic ages ranges of
  0--100\,Myr and 49--51\,Myr, and radial velocity distributions
  varying from a uniform distribution (solid line) to Gaussian
  distributions with velocity dispersions of 60\,km\,s$^{-1}$
  (long-dashed line), 130\,km\,s$^{-1}$ (short-dashed line), and
  200\,km\,s$^{-1}$ (dash-dotted line). For clarity, the PDFs are
  offset from each other by an arbitrary amount.}
\label{vcm}
\end{figure*}
 
Since the post-SN peculiar velocity represents one of the most
important elements in constraining the kick velocity imparted to
pulsar~B at birth, it is worth taking a closer look at the range and
distribution of the post-SN peculiar velocities derived from the
Galactic motion calculations. In Fig.~\ref{vcm}, the distribution of
post-SN peculiar velocities is shown for a present-day transverse
velocity component of 30\,km\,s$^{-1}$, and kinematic age ranges of
0-100\,Myr and 49-51\,Myr. These probability distributions are
calculated considering weights according to (i) the probability that
the system has a present-day radial velocity $V_r$ (assumed to be
distributed according to a uniform or a Gaussian distribution), (ii)
the transverse velocity has a direction angle $\Omega$ (assumed to be
uniformly distributed), and (iii) the system is found at its current
position in the Galaxy (determined by the time the system spends near
its current position divided by its kinematic age, see \S\,\ref{pos}
for details).  In particular, the probability that, for a given pair of $V_r$ and $\Omega$, the post-SN peculiar velocity of the binary is equal to $V_{\rm CM}$ is given by
\begin{equation}
P \left( V_{\rm CM}| V_r, \Omega \right) \propto 
  \sum_{i=1}^{N(V_r,\Omega)} {{ T \left( V_r, \Omega \right) } 
  \over {\tau_{{\rm kin},i} \left( V_r, \Omega \right)}}\, 
  \lambda_i \left( V_r, \Omega \right),  \label{pvcm}
\end{equation}
where $N(V_r,\Omega)$ is the number of Galactic plane crossings associated with $V_r$ and $\Omega$, $T(V_r, \Omega)$ is the time the system spends near its current position for the orbit associated with $V_r$ and $\Omega$, and $\tau_{{\rm kin},i} \left( V_r, \Omega \right)$ is the kinematic age corresponding to the $i$-th plane crossing associated with $V_r$ and $\Omega$. The factor $\lambda_i \left( V_r, \Omega \right)$ is equal to 1 when the $i$-th plane crossing along the orbit associated with $V_r$ and $\Omega$ yields a post-SN peculiar velocity equal to $V_{\rm CM}$, and equal to 0 otherwise. The total probability that the post-SN peculiar velocity of the binary is equal to $V_{\rm CM}$ is then obtained by integrating Eq.~(\ref{pvcm}) over all possible values of $V_r$ and $\Omega$:
\begin{equation}
P \left( V_{\rm CM} \right) \propto \int_{\Omega} \int_{V_r} 
  P \left( V_{\rm CM}| V_r, \Omega \right) P \left( V_r \right)
  P \left( \Omega \right) dV_r\, d\Omega.  \label{pvcm2}
\end{equation}
Here $P(V_r)$ is the probability distribution of the unknown radial velocity $V_r$, and $P(\Omega)$ the probability distribution of the unknown proper motion direction $\Omega$ in the plane perpendicular to the line-of-sight. 

For ages of 0-100\,Myr, post-SN peculiar velocities up to
100\,km\,s$^{-1}$ are likely for all four radial velocity
distributions. The highest post-SN peculiar velocities are found for
the Gaussian radial velocity distributions with velocity dispersions
of 130\,km\,s$^{-1}$ and 200\,km\,s$^{-1}$ ($V_{\rm CM}$ values remain
likely up to 200--300\,km\,s$^{-1}$) and the uniform radial velocity
distribution ($V_{\rm CM}$ values follow an almost flat distribution
from 300\,km\,s$^{-1}$ to 1500\,km\,s$^{-1}$). For ages of 49-51\,Myr,
the post-SN peculiar velocity distributions all have a peak at
50\,km\,s$^{-1}$. In the case of the uniform radial velocity
distribution and the Gaussian distributions with velocity dispersions
of 130\,km\,s$^{-1}$ or 200\,km\,s$^{-1}$, additional peaks of almost
equal height occur at even higher post-SN peculiar velocities. Similar
conclusions apply to the post-SN peculiar velocity distributions
obtained for a present-day transverse velocity component of
10\,km\,s$^{-1}$.  Despite the small lower limit on the present-day
{\em transverse} systemic velocity, a wide range of non-negligible post-SN
peculiar velocities is thus possible. We note however that the
distributions incorporate all possible post-SN peculiar velocities
obtained from tracing the motion of the system backwards in time and
that some of these may require SN kicks and mass loss that are
incompatible with the observational constraints on the orbital
elements and component masses of the double pulsar.

\begin{figure*}
\resizebox{14.5cm}{!}{\includegraphics{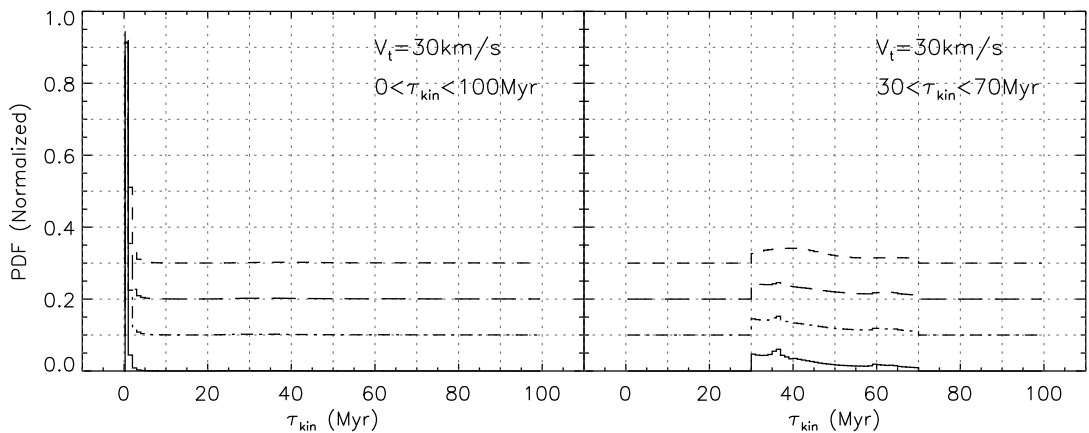}}
\caption{Distribution of kinematic ages for a present-day transverse
  velocity of 30\,km\,s$^{-1}$, kinematic age ranges of 0--100\,Myr
  and 30--70\,Myr, and radial velocity distributions varying from a
  uniform distribution (solid line) to Gaussian distributions with
  velocity dispersions of 60\,km\,s$^{-1}$ (long-dashed line),
  130\,km\,s$^{-1}$ (short-dashed line), and 200\,km\,s$^{-1}$
  (dash-dotted line). For clarity, the PDFs are offset from each other
  by an arbitrary amount.}
\label{tauk}
\end{figure*}

Similarly, PDFs can be constructed for the kinematic ages $\tau_{\rm
kin}$. They are shown in Fig.~\ref{tauk} for a present-day transverse
velocity component of 30\,km\,s$^{-1}$, and kinematic age ranges of
0-100\,Myr and 30-70\,Myr. For ages of 0--100\,Myr, the PDFs are all
very strongly peaked at very young ages. In particular, 90--97\% of
the Galactic plane crossings give rise to ages of 0--5\,Myr. The peak
of the PDF at ages of 0--5\,Myr furthermore increases with increasing
velocity dispersion of the adopted radial velocity distribution. For
ages of 30--70\,Myr, the PDFs overall decrease with increasing values
of $\tau_{\rm kin}$ and show no strong preference for any particular
range of kinematic age values. Similar conclusions apply to the
kinematic ages obtained for a present-day transverse velocity
component of 10\,km\,s$^{-1}$.

\subsection{Theoretical radial velocity distributions}
\label{pvrad}

\begin{figure*}
\resizebox{14.5cm}{!}{\includegraphics{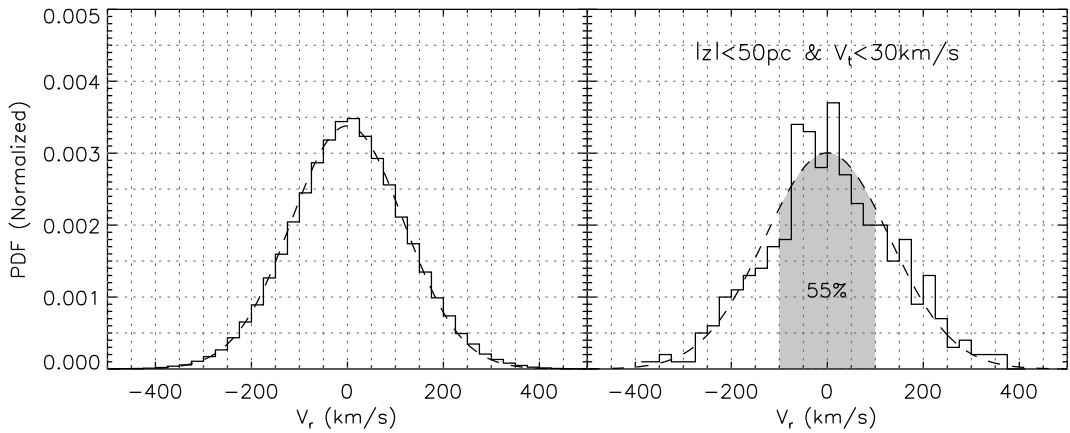}}
\caption{Present-day radial-velocity distributions of coalescing DNSs
  for a population synthesis model with Maxwellian kicks with a
  velocity dispersion $\sigma_{\rm kick}=250\,{\rm km\,s^{-1}}$. The
  left panel shows the distribution for the entire population of
  coalescing DNSs, while the right panel shows the distribution of
  coalescing DNSs within 50\,pc of the Galactic mid-plane and with a
  proper motion of less than 30\,km\,s$^{-1}$. The dashed lines in the
  left and right panels are best-fit Gaussian distributions with
  velocity dispersions of 117\,km\,s$^{-1}$ and 130\,km\,s$^{-1}$,
  respectively. Note that in the right panel 45\% of the systems have
  $|V_r| > 100\,{\rm km\,s^{-1}}$.}
\label{vr250}
\end{figure*}

\begin{table}
\caption{Velocity Dispersions for Supernova Kicks and \\ Radial Velocities
\label{fits}}
\begin{ruledtabular}
\begin{tabular}{cc}
$\sigma_{\rm kick}$\footnote{Maxwellian distribution.} & 
$\sigma_{V_r}$\footnote{Best fit Gaussian distribution centered at
    zero. Values are for coalescing DNSs within 50\,pc of the Galactic
    mid-plane and with a  proper motion of less than
    30\,km\,s$^{-1}$.} \\
(km\,s$^{-1}$)        & (km\,s$^{-1}$) \\
\hline \hline
 50 &  $63 \pm 2$ \\
100 &  $94 \pm 4$ \\
150 &  $92 \pm 4$ \\
200 & $104 \pm 5$ \\
250 & $130 \pm 7$ \\
300 & $148 \pm 9$ \\
350 & $124 \pm 7$ \\
400 & $172 \pm 15$ \\
450 & $147 \pm 14$ \\
500 & $280 \pm 141$ \\ 
\end{tabular}
\end{ruledtabular}
\end{table}

While the radial velocity of PSR\,J0737-3039 is unknown, population
synthesis calculations can provide us with some reasonable guidance on
the expected radial velocity distribution of the DNS population.  We
use the {\em StarTrack} population synthesis code (see
\citet{2002ApJ...572..407B}; and especially
\citet{2005astro.ph.11811B} for an extensive and detailed description
of this binary population synthesis evolutionary code) to develop
models of DNS formation and evolution in our Galaxy (assuming
continuous star-formation history over 10\,Gyr and solar
metallicity). To calculate the DNS kinematic evolution we need to also
choose a spatial distribution of DNS binary progenitors in the
Galaxy. We adopt a standard double exponential to represent the
density distribution of Galactic systems:
\begin{equation}
n(R,z) = n_o\, \exp \left( -R/H_R \right) 
  \exp \left( -|z|/H_Z \right),  \label{nrz}
\end{equation}
where $n_o = 1/(4 \pi H_Z H_R^2)$ is the normalization, $H_R$ the
scale length, and $H_Z$ the scale height. We adopt $H_R = 2.8$\,kpc
and $H_Z = 70$\,pc, appropriate for massive stars. The motion of the
DNSs in the Galaxy is calculated using the same Galactic potential as
in \S\,\ref{motion}.

In this analysis, we consider a set of models with input parameters
chosen as for the standard model (model A) in
\citet{2002ApJ...572..407B}, except for the assumed Maxwellian
distributions of the kick magnitudes imparted to nascent neutron
stars: we vary the one-dimensional kick velocity dispersion
$\sigma_{\rm kick}$ in the range of $50-500$\,km\,s$^{-1}$, as shown
in Table\,\ref{fits}. We restrict our parameter study to these models,
because a broad model exploration is not the primary goal of our study
and because prior experience with DNS models clearly indicates that
the assumed NS kick velocity model is the most important parameter
affecting the DNS kinematic evolution and properties
\citep{2002ApJ...571L.147B}.

We use this set of DNS models to derive radial-velocity distributions
with respect to the Sun. To obtain a realistic representation of the
kinematic properties of the {\em current} population of tight DNS
systems (like PSR\,J0737-3039), we place a cut on the model DNS
populations and select only those that have already formed (i.e., they
have a formation time shorter than 10\,Gyr) and will coalesce within
the age of the Universe (at a time in the range 9--14\,Gyr). We next
distribute the selected DNS population in the Galaxy and follow their
motion using the calculated center-of-mass velocities imparted to the
binaries after both of the two supernova events \footnote{In kinematic
calculations of DNS binaries the part of the motion preceding the
second supernova explosion is most often ignored as
negligible. However, we have examined our radial velocity
distributions with and without this early motion and we can confirm
that its effect is indeed negligible. This is not so much due to the
somewhat smaller center-of-mass velocities after the first explosion
(compared to those after the second explosion because of the wide
orbits involved); instead it is mostly due to the very short time
between between the two events (from about one tenth to a few
Myr).}. We also take into account the effect of Galactic rotation. We
then derive the radial velocity distribution by extracting the
calculated peculiar velocities (relative to the Sun) and considering
the appropriate projections. We can further impose a number of
additional constraints on the model population to select systems that
have kinematic properties similar to PSR\,J0737-3039: coalescing DNS
with proper motion smaller than 30\,km\,s$^{-1}$ and a location within
50\,pc from the Galactic plane; we note that we do not impose any
constraints on the binary properties to avoid any statistical accuracy
problems.  

We find that the derived radial velocity distributions are very well
approximated by a Gaussian centered at zero with velocity dispersions
$\sigma_{\rm V_r}$ ranging from 60 to 200\,km\,s$^{-1}$. The
dependence of the derived radial-velocity dispersions on the kick
velocity dispersions are shown in Table~\ref{fits}.  The distributions
with and without the kinematic constraints are shown in
Figure\,\ref{vr250}, for $\sigma_{\rm kick}=250\,{\rm
km\,s^{-1}}$. The corresponding values of $\sigma_{\rm V_r}$ for the
two panels are rather similar: 117\,km\,s$^{-1}$ and
130\,km\,s$^{-1}$, respectively.  Moreover, for this $\sigma_{\rm
kick}$ value, 45\% of the systems have $|V_r| > 100\,{\rm
km\,s^{-1}}$.

It is important to note that imposing the proper-motion constraint
does not actually reduce the typical magnitude of the radial
velocity. The reason is that small proper motion values select two
sub-populations of DNS systems: one with inherently small
center-of-mass velocities (and consequently small radial velocities),
and another with large center-of-mass velocities but appropriate
direction so that the proper motion is small. This second group
naturally leads to large radial velocities because, for a given
center-of-mass velocity, the projection affects the radial component
in the opposite way as it does the transverse component.  \citet{2006astro.ph..3649P} somewhat misleadingly draw the focus away from this second group and use the first sub-population to argue that both the present {\em and} post-SN peculiar velocity must be small. Since our theoretical radial-velocity distributions incorporate the constraints $|Z|<50$\,pc and $V_t < 30\,{\rm km\,s^{-1}}$, they implicitely account for the geometrical probability argument by \citet{2006astro.ph..3649P} that the probability of a small transverse velocity component is larger for smaller total systemic peculiar velocities (see their Eq. 2). Moreover, as outlined in detail in the previous section, we explicitely account for differences between the system's current and post-SN peculiar velocity by modeling its past motion in the Galactic potential.

Last, we note (although not directly relevant to this study) that with
the full population synthesis models we confirm the results of
\citet{2001ApJ...556..340K} about the vertical DNS distribution: for
$\sigma_{\rm kick}=250\,{\rm km\,s^{-1}}$, it is best fitted by two
exponentials, each with a typical scale height of $0.230 \pm
0.006$\,kpc and $1.72 \pm 0.07$\,kpc. As expected the scale height
increases with increasing typical kick magnitude.

\subsection{Probability to find the system at its current position}
\label{pos}

At a distance of $\sim 600$\,pc and a Galactic latitude of
$-4.5^\circ$, PSR\,J0737-3039 is currently located only 20\,pc below
the Galactic plane. \citet{2005PhRvL..94e1102P} used the proximity of
the system to the Galactic plane to argue that pulsar~B most likely
received only a small kick velocity at birth and that its helium star
progenitor most likely had a pre-SN mass only slightly larger than the
pulsar's present mass.  Their procedure, however, only accounted for
the motion of the system perpendicular to the Galactic plane.

Here, we determine the probability of finding the system at its
current position in the Galaxy in three dimensions (i.e., besides the
vertical distance to the Galactic plane, we also consider the radial
and azimuthal position in the plane). For this purpose, we determine
the time the system spends in a sphere with radius $R$ centered on its
current location for all $V_r$- and $\Omega$-values considered for the
derivation of the possible birth sites in \S\,\ref{motion}. The probability to find the
system near its current position is then determined as the time it
spends in this sphere divided by its age. For the latter, we adopt the
kinematic ages obtained by tracing the motion of the system backwards
in time, so that the probability of finding the system near its current
position depends on the radial velocity $V_r$, the proper motion
direction $\Omega$, and the Galactic plane crossing considered along
the orbit associated with $V_r$ and $\Omega$.  We performed some test
calculations to successfully verify that the results based on the thus
determined probabilities are insensitive to the adopted value of $R$,
as long as it is sufficiently small for the sphere to represent a
local neighborhood near the system's current position. For the results
presented in this paper we, somewhat arbitrarily, choose $R=50$\,pc.

\section{Progenitor constraints}
\label{progenitor}

\subsection{Post-supernova orbital parameters}

After the formation of pulsar~B, the evolution of the system is
expected to be driven exclusively by the emission of gravitational
waves. The post-SN orbital semi-major axis and eccentricity can then
obtained from the current values by numerically integrating the system
of differential equations governing the orbital evolution due to
gravitational wave emission backwards in time. For this purpose, we
use the differential equations derived by \citet{1992MNRAS.254..146J}
which are valid up to 3.5 post-Newtonian order of approximation.

We calculate the orbital evolution backwards in time for all Galactic
plane crossings associated with all values of $V_r$ and $\Omega$
considered in the calculation of the possible birth sites. For each
crossing the integration is terminated at the associated kinematic age
$\tau_{\rm kin}$. We find the resulting post-SN orbital separation $A$
to be always between $1.26\, R_\odot$ and $1.54\, R_\odot$, and the
post-SN orbital eccentricity $e$ between $0.088$ and $0.12$ (see
Papers~I and~II). Since the kinematic ages are functions of the radial
velocity $V_r$, the proper motion direction $\Omega$, and the Galactic
plane crossing considered along the past orbit associated with $V_r$
and $\Omega$, the particular values of $A$ and $e$ associated with a
given $\tau_{\rm kin}$ are also functions of these variables.

\subsection{Orbital dynamics of asymmetric supernova explosions}
\label{asymSN}

The pre- and post-SN binary parameters and the kick velocity imparted
to pulsar~B at birth are related by the conservation laws of orbital
energy and angular momentum. For a circular pre-SN orbit, the
relations take the form \citep[e.g.][]{1983ApJ...267..322H,
1995MNRAS.274..461B, 1996ApJ...471..352K, 1996ApJ...471..352K,
1997ApJ...489..244F, 2000ApJ...530..890K}
\begin{eqnarray}
\lefteqn{V_{\rm k}^2 + V_0^2 
 + 2\, V_{\rm k}\, V_0\, \cos
 \theta} \nonumber \\
 & & = G \left( M_A + M_B \right) \left( {2 \over A_0}
 - {1 \over A} \right),  \label{eq1}
\end{eqnarray}
\begin{eqnarray}
\lefteqn{A_0^2 \left[ V_{\rm k}^2\, \sin^2 \theta \cos^2
 \phi \right. + \left. \left( V_{\rm k}\, \cos \theta 
 + V_0 \right)^2 \right]} \nonumber \\
 & = & G \left( M_A + M_B \right) A 
 \left( 1 - e^2 \right), \hspace{1.2cm} \label{eq2}
\end{eqnarray}
where $M_0$ is the mass of pulsar~B's pre-SN helium star progenitor,
$A_0$ the pre-SN orbital separation, $V_0=[G(M_A+M_0)/A_0]^{1/2}$ the
relative orbital velocity of pulsar~B's pre-SN helium star progenitor,
and $V_k$ the magnitude of the kick velocity imparted to pulsar~B at
birth. The angles $\theta$ and $\phi$ define the direction of the kick
velocity imparted to pulsar~B: $\theta$ is the polar angle between
$\vec{V}_k$ and $\vec{V}_0$, and $\phi$ the corresponding azimuthal
angle in the plane perpendicular to $\vec{V}_0$ defined so that
$\phi=\pi/2$ corresponds to the direction from pulsar~A to pulsar~B's
helium star progenitor (see Fig.~1 in \citep{2000ApJ...541..319K} for
a graphical representation).

The requirement that Eqs.~(\ref{eq1}) and~(\ref{eq2}) permit real
solutions for $V_k$, $\theta$, $\phi$, $M_0$, and $A_0$, imposes
constraints on the pre- and post-SN binary parameters and on the
magnitude and direction of the kick velocity imparted to pulsar~B at
birth. For a mathematical formulation of these constraints, we refer
to Eqs.~(21)--(27) in \citep{2005ApJ...625..324W}, and references
therein (a more compact description can also be found in Papers~I
and~II). We here merely recall that the constraints express that: (i)
the binary components must remain bound after the SN explosion, (ii)
the pre- and post-SN orbits must pass through the instantaneous
position of the component stars at the time of the SN explosion, and
(iii) there is a lower and upper limit on the degree of orbital
contraction or expansion associated with a given amount of mass loss
and a given SN kick.

Besides the changes in the orbital parameters and the mass of the
exploding star, the SN explosion also imparts a kick velocity to the
binary's center of mass and tilts the post-SN orbital plane with
respect to the pre-SN one. Under the assumption that the pre-SN
peculiar velocity of the binary is small in comparison to the pre-SN
orbital velocity of the component stars, the magnitude of the post-SN
center-of-mass velocity $V_{\rm CM}$ is given by
\citep[e.g.][]{1996ApJ...471..352K} 
\begin{eqnarray}
\lefteqn{V_{\rm CM}^2 = {{M_B\,M_0} \over {(M_A+M_0)
  (M_A+M_B)}} \left[ {{G (M_0-M_B) M_A} \over {M_0\, A}}
  \right. }  \nonumber \\
 & & \left. + {{G (M_0-M_B) (M_0-2\,M_B) M_A} \over 
  {M_0\, M_B\, A_0}} + V_k^2  \right].
  \hspace{1.7cm} \label{vpec}
\end{eqnarray}
The tilt angle $\lambda$ between the pre- and post-SN orbital planes 
is given by \citep{2000ApJ...541..319K}
\begin{eqnarray}
\lefteqn{ \cos \lambda = \left[ 
  {A \over A_0}\, 
  {{M_A + M_B} \over {M_A + M_0}}
  \left( 1-e^2 \right) \right]^{-1/2} } \nonumber \\
 & & \times \left( {V_k \over V_0}\, 
  \cos \theta + 1 \right). \hspace{3.0cm} \label{tilt}
\end{eqnarray}
From Eqs.~(\ref{eq2}) and~(\ref{tilt}), one derives that $\lambda =
0^\circ$ for kicks directed in the pre-SN orbital plane (i.e. $\cos
\theta = \pm 1$), independent of the kick velocity magnitude $V_k$. 

\subsection{Stellar and binary evolution}
\label{evol}

The constraints on the progenitor of PSR\,J0737-3039 resulting from
the dynamics of asymmetric SN explosions arise solely from tracing the
evolution of the current system properties backwards in time. The
pre-SN orbital separations and pulsar~B progenitor masses found this
way are, however, not necessarily accessible through the currently
known DNS formation channels. Further constraints on the progenitor of
pulsar~B can therefore be obtained from stellar and binary evolution
calculations.

A lower limit on the mass of pulsar~B's pre-SN progenitor is given by
the requirement that the star must be massive enough to evolve into a
NS rather than a white dwarf. According to our current understanding
of helium star evolution, the minimum helium star mass required for NS
formation is 2.1--2.8\,$M_\odot$ \citep{1986A&A...167...61H,
2003astro.ph..3456T}. The actual helium star minimum mass is, however,
still considerably uncertain due to the poorly understood evolution of
massive stars and possible interactions with close binary companions.

As in Papers~I and~II, we impose a lower limit of $2.1\,M_\odot$ on
the mass of pulsar~B's pre-SN helium star progenitor. However, in the
light of recent suggestions that the progenitor of pulsar~B may have
been significantly less massive than the conventional lower limit of
$2.1\,M_\odot$, we also explore the possibility of progenitor masses
as low as pulsar B's present-day mass of $1.25\,M_\odot$
\citep{2005PhRvL..94e1102P, 2005MNRAS.361.1243P}. Unless our current
understanding of helium star evolution is seriously flawed, this
scenario implies that the progenitor of pulsar~B must have lost a
significant amount of mass (at least $\sim 0.7\,M_\odot$) after it had
already established a sufficiently massive core to guarantee the
occurrence of a future SN explosion.  We note that this possibility is included in the binary population synthesis calculations used to derive the adopted theoretical DNS radial-velocity distributions (see Table~\ref{fits}). Contrary to what is stated by \citet{2006astro.ph..3649P}, our population synthesis calculations do allow for low-mass helium-rich progenitors of the second NS and the derived radial-velocity distributions are therefore by no means a priori biased towards high progenitor masses.

In Paper~I, we also showed that the pre-SN binary was so tight
($1.2\,R_\odot \lesssim A_0 \lesssim 1.7\,R_\odot$) that the helium
star progenitor of pulsar~B must have been overflowing its Roche lobe
at the time of its SN explosion (see also
\citep{2004MNRAS.349..169D}). An upper limit on the progenitor mass is
therefore given by the requirement that this mass-transfer phase be
dynamically stable (otherwise the components would have merged and no
DNS would have formed)\footnote{An updated population synthesis study on coalescing DNSs shows that some systems can actually survive the dynamically unstable mass-transfer phase and form a DNS  (Belczynski et al., in preparation). In view of our conservative upper limit of $4.7\,M_\odot$ on the mass of pulsar~B's pre-SN helium star progenitor and the small values of our PDFs for masses near this upper limit, we do not expect this possibility to significantly affect the results presented in this paper.}. Based on the evolutionary tracks for NS +
helium star binaries calculated by \citet{2003ApJ...592..475I}, we
adopt an upper limit of 3.5 on the mass ratio $M_0/M_A$ of the pre-SN
binary to separate dynamically stable from dynamically unstable
Roche-lobe overflow (see also \citep{2002MNRAS.331.1027D,
2003MNRAS.344..629D}). For $M_A=1.34\,M_\odot$, this yields an upper
limit of $4.7\,M_\odot$ on the mass of pulsar~B's pre-SN helium star
progenitor. Because of the mass transfer, the pre-SN progenitor mass
may, however, be considerably lower than the progenitor mass at the
onset of Roche-lobe overflow. The upper limit of $4.7\,M_\odot$ is
therefore fairly conservative. However, as we will see in the next
section, the preferred pre-SN progenitor masses are always
significantly smaller than $4.7\,M_\odot$ so that a stricter upper
limit would not considerably affect any of our results.

\section{Probability distribution functions}
\label{results}

\subsection{A coherent picture for the progenitor and formation of
  pulsar~B}  

The steps outlined in the previous sections have been used to derive
the most likely pre-SN progenitor of PSR\,J0737-3039 and the magnitude
and direction of the kick velocity imparted to pulsar~B at
birth. Firstly, the possible kinematic ages and post-SN peculiar
velocities are determined by tracing the Galactic motion backwards in
time using the current constraints on the proper motion and position
in the Galaxy. These ages and velocities are functions of the
presently unknown values of the radial velocity $V_r$ and proper
motion direction $\Omega$. Secondly, the kinematic ages are used to
reverse the orbital evolution due to gravitational radiation and
determine the orbital semi-major axis and eccentricity right after the
formation of pulsar~B. Thirdly, the pre-SN binary parameters and kick
velocity imparted to pulsar~B are constrained using the conservation
of orbital energy and angular momentum during asymmetric SN
explosions. The constraints are supplemented with the requirements
that the post-SN peculiar velocity obtained from Eq.~(\ref{vpec}) be
compatible with the post-SN peculiar velocity obtained by tracing the
motion of the system in the Galaxy backwards in time, and that the
spin-orbit misalignment of pulsar~A obtained from Eq.~(\ref{tilt}) be
compatible with the observational constraint $|\lambda - 90^\circ| >
30^\circ$. Imposing the latter constraint rests on the additional
assumption that pulsar~A's spin is aligned with the pre-SN orbital
angular momentum, as expected from the strong tidal interactions
operating during the spin-up phase of pulsar~A.

\subsection{The most likely kick velocity and progenitor properties}
\label{mostlikely}

Since the conservation laws of orbital energy and angular momentum
yield two equations for five unknown quantities $V_k$, $\cos \theta$,
$\phi$, $M_0$, and $A_0$, we solve the equations for $M_0$ and $A_0$
as functions of $V_k$, $\cos \theta$, and $\phi$. Each solution is then
weighted assuming uniform prior distributions for $V_k$, $\cos \theta$,
$\phi$, $M_0$, and $A_0$. Thus the kick direction is assumed to be isotropically distributed in space. 
Similar to Eqs.~(\ref{pvcm}) and (\ref{pvcm2}), we next determine the probability that, given the currently known observational constraints, PSR\,J0737-3039B was born with a natal kick velocity $V_k$ with a direction determined by $\cos \theta$ and $\phi$ from a progenitor of mass $M_0$ orbiting the first-born pulsar in a circular orbit of radius $A_0$ as
\begin{eqnarray}
\lefteqn{P \left( V_k, \cos \theta, \phi, M_0, A_0 | 
  X_{\rm obs} \right)} \nonumber \\
 & & \propto \int_{\Omega} \int_{V_r} \left[
  \sum_{i=1}^{N(V_r,\Omega)} {{ T \left( V_r, \Omega \right) } 
  \over {\tau_{{\rm kin},i} \left( V_r, \Omega \right)}}\, 
  \kappa_i \left( V_r, \Omega \right) \right]  \label{pprg2} \\
 & & \times P \left( V_r \right) P \left( \Omega \right) 
  dV_r\, d\Omega. \nonumber
\end{eqnarray}
Here $X_{\rm obs}$ denotes all currently known observational constraints (orbital separation, eccentricity, proper motion, etc). The factor $\kappa_i \left( V_r, \Omega \right)$ is equal to 1 when the considered $V_k$, $\cos \theta$, $\phi$, $M_0$, and $A_0$ satisfy all constraints imposed by tracing the system's kinematic and evolutionary history back to the formation time of pulsar~B for the $i$-th plane crossing along the orbit in the Galaxy associated with $V_r$ and $\Omega$, and equal to 0 otherwise. The other quantities are defined in \S\,\ref{motion}. Hence, we obtain a 5-dimensional PDF for the kick velocity and progenitor parameters: $V_k$, $\cos \theta$, $\phi$, $M_0$, and $A_0$.

\begin{table}
\caption{Effect of marginalizing the 5-D PDF for $V_k$, $\cos \theta$,
  $\phi$, $M_0$, and $A_0$ on the most likely kick velocity and
  progenitor parameters for a transverse velocity component of
  $30\,{\rm km\,s^{-1}}$, a minimum helium star mass of
  $1.25\,M_\odot$, an age range of 0--100\,Myr, and a uniform
  present-day radial velocity distribution.
\label{marg}}
\begin{ruledtabular}
\begin{tabular}{lccccc}
          & \multicolumn{5}{c}{\textbf{Most likely value}} \\  \cline{2-6}
\textbf{Variables} & \textbf{$V_k$} & \textbf{$\cos \theta$} & \textbf{$\phi$} & \textbf{$M_0$} & \textbf{$A_0$} \\
   & (km\,s$^{-1}$) &       & (deg) & ($M_\odot$) & ($R_\odot$) \\
\hline
$V_k$, $\cos \theta$, $\phi$, $M_0$, $A_0$ & 75 & -0.025 & 10 & 1.45 & 1.15 \\
$V_k$, $\cos \theta$, $\phi$, $M_0$        & 75 & -0.025 & 10 & 1.45 &      \\
$V_k$, $\cos \theta$, $M_0$                & 45 & -0.050 &    & 1.45 &      \\
$V_k$, $M_0$                          & 10 &        &    & 1.45 &      \\
$V_k$                                 & 55 &        &    &      &      \\
\end{tabular}
\end{ruledtabular}
\end{table}

The construction of the full 5-D PDF represents a significant
departure from previous investigations which restricted the analysis
to the derivation of 1-D PDFs for $V_k$ (Papers~I and~II) or 2-D PDFs
for $V_k$ and $M_0$ \citep{2005PhRvL..94e1102P,
2005astro.ph.10584P}. Although the derivation of the most likely $V_k$
and $M_0$ values from these marginalized PDFs is mathematically
correct, they may be subjected to projection effects which can
drastically affect the position of the maxima. In order to illustrate
this, we constructed the 5-D PDF for a transverse velocity component
of $30\,{\rm km\,s^{-1}}$, a minimum helium star mass of
$1.25\,M_\odot$, an age range of 0--100\,Myr, and a uniform
present-day radial velocity distribution. The most likely kick
velocity and progenitor properties obtained from the 5-D PDF are $V_k
\simeq 75\,{\rm km\,s^{-1}}$, $\cos \theta \simeq -0.025$, $\phi =
10^\circ$, $M_0 \simeq 1.45\,M_\odot$, and $A_0 \simeq
1.15\,R_\odot$. The most likely values obtained after successively
integrating over $A_0$, $\phi$, $\cos \theta$, and $M_0$ are listed in
Table~\ref{marg}. While the integration over $A_0$ does not
significantly affect the most likely value of $V_k$, successive
integrations over $\phi$ and $\cos \theta$ reduce the most likely
$V_k$ to 45\,km\,s$^{-1}$ and 10\,km\,s$^{-1}$, respectively. A
subsequent integration over $M_0$ furthermore increases the most
likely $V_k$ back to 55\,km\,s$^{-1}$. Marginalizing a
multi-dimensional PDF can thus have a significant impact on the
location of its peaks. Some caution is therefore in order when
interpreting marginalized PDFs.

\squeezetable\begin{table*}
\caption{The most likely kick velocity and progenitor parameters
  obtained from the 5-D PDF for $V_k$, $\cos \theta$, $\phi$, $M_0$,
  and $A_0$, for different transverse systemic velocity components,
  minimum pre-SN helium star masses, DNS ages, and present-day radial
  velocity distributions. For brevity, the uniform radial velocity
  distribution is denoted by $\sigma_{V_r}=\infty$. Tabulated values
  of $\phi$ are restricted to $0^\circ \le \phi \le 90^\circ$. For
  each listed $\phi$-value additional and equally likely solutions to
  Eqs.~(\ref{eq1}) and~(\ref{eq2}) are associated with $-\phi$ and
  $180^\circ \pm \phi$. For $V_k$ and $M_0$, 68\% and 90\% confidence
  intervals are also listed. The intervals are obtained by
  marginalizing the 5-D PDF to 1-D PDFs for $V_k$ and $M_0$.
\label{pdfmax}}
\begin{ruledtabular}
\begin{tabular}{cccccccccccccccc}
      \multicolumn{4}{c}{\textbf{Priors}} & \phantom{x} & 
      \multicolumn{5}{c}{\textbf{Most likely progenitor}} & \phantom{x} &
      \multicolumn{2}{c}{\textbf{68\% confidence intervals}} & \phantom{x} & 
      \multicolumn{2}{c}{\textbf{90\% confidence intervals}} \\
      \cline{1-4} \cline{6-10} \cline{12-13} \cline{15-16}
\textbf{$V_t$} & \textbf{$M_{\rm 0,min}$} & \textbf{$\tau_{\rm kin}$} & \textbf{$\sigma_{V_r}$} &  & \textbf{$V_k$} & \textbf{$\cos \theta$} & \textbf{$\phi$} & \textbf{$M_0$} & \textbf{$A_0$} &  
   & \textbf{$V_k$} & \textbf{$M_0$} & & \textbf{$V_k$} & \textbf{$M_0$}  \\
 (km\,s$^{-1}$) & ($M_\odot$) & (Myr) & (km\,s$^{-1}$) &  & (km\,s$^{-1}$) & (deg) & (deg) & ($M_\odot$) & ($R_\odot$) &  
   & (km\,s$^{-1}$) & ($M_\odot$) &  & (km\,s$^{-1}$) & ($M_\odot$) \\
\hline
10 &  2.1 & 0-100 & $\infty$ & & 170 & -0.925 & 45 & 2.65 & 1.35 & & 105-335 &  2.1-3.3 & & 75-535 &  2.1-4.2 \\
10 &  2.1 & 0-100 & 200 & & 160 & -0.925 & 45 & 2.55 & 1.35 & &  95-255 &  2.1-2.8 & & 75-345 &  2.1-3.5 \\
10 &  2.1 & 0-100 & 130 & & 125 & -0.900 & 65 & 2.25 & 1.35 & &  95-215 &  2.1-2.6 & & 75-285 &  2.1-3.1 \\
10 &  2.1 & 0-100 & 60  & &  80 & -0.950 & 15 & 2.15 & 1.15 & &  75-155 &  2.1-2.4 & & 65-195 &  2.1-2.6 \\

10 &  2.1 & 30-70 & $\infty$ & & 120 & -0.875 & 75 & 2.25 & 1.45 & &  95-285 &  2.1-3.1 & & 75-405 &  2.1-3.9 \\
10 &  2.1 & 30-70 & 200 & & 120 & -0.875 & 80 & 2.25 & 1.45 & &  95-255 &  2.1-2.9 & & 65-345 &  2.1-3.6 \\
10 &  2.1 & 30-70 & 130 & & 120 & -0.875 & 80 & 2.25 & 1.45 & &  85-225 &  2.1-2.7 & & 65-305 &  2.1-3.3 \\
10 &  2.1 & 30-70 & 60  & &  75 & -0.950 & 15 & 2.15 & 1.25 & &  75-155 &  2.1-2.4 & & 65-205 &  2.1-2.6 \\

10 &  2.1 & 49-51 & $\infty$ & &  70 & -0.950 & 15 & 2.15 & 1.25 & & 65-105, 165-355 & 2.1-2.3, 2.7-3.8 & & 65-455 &  2.1-4.1 \\
10 &  2.1 & 49-51 & 200 & &  70 & -0.950 & 15 & 2.15 & 1.25 & & 65-125, 165-315 &  2.1-2.4, 2.7-3.6 & & 65-395 &  2.1-3.8 \\
10 &  2.1 & 49-51 & 130 & &  70 & -0.950 & 15 & 2.15 & 1.25 & & 65-135, 165-265 &  2.1-2.4, 2.8-3.3 & & 65-345 &  2.1-3.3, 3.5-3.7 \\
10 &  2.1 & 49-51 & 60  & &  70 & -0.950 & 15 & 2.15 & 1.25 & &  65-125 &  2.1-2.3 & & 65-215 &  2.1-2.8  \\

10 & 1.25 & 0-100 & $\infty$ & & 170 & -0.925 & 45 & 2.65 & 1.35 & &   0-275 &  1.3-2.8 & &  0-475 &  1.25-3.8 \\
10 & 1.25 & 0-100 & 200 & & 160 & -0.925 & 45 & 2.55 & 1.35 & &   0-185 &  1.3-2.3 & &  0-305 &  1.25-2.9 \\
10 & 1.25 & 0-100 & 130 & &  50 & -0.050 & 75 & 1.35 & 1.25 & &   0-135 & 1.25-1.9 & &  0-235 &  1.25-2.5 \\
10 & 1.25 & 0-100 & 60  & &  50 & -0.050 & 75 & 1.35 & 1.25 & &   0-65  &  1.3-1.6 & &  0-115 &  1.25-1.8 \\

10 & 1.25 & 30-70 & $\infty$ & &   5 & -0.200 & 20 & 1.55 & 1.25 & &   0-145 & 1.25-2.0 & &  0-295 &  1.25-3.0 \\
10 & 1.25 & 30-70 & 200 & &   5 & -0.175 & 15 & 1.55 & 1.25 & &   0-115 & 1.25-1.8 & &  0-235 &  1.25-2.6 \\
10 & 1.25 & 30-70 & 130 & &   5 & -0.175 & 15 & 1.55 & 1.25 & &   0-95  & 1.25-1.7 & &  0-195 &  1.25-2.3 \\
10 & 1.25 & 30-70 & 60  & &   5 & -0.175 & 15 & 1.55 & 1.25 & &   0-65  &  1.3-1.6 & &  0-105 &  1.25-1.8 \\

10 & 1.25 & 49-51 & $\infty$ & &  95 & -0.775 & 80 & 1.95 & 1.45 & &   0-155 &  1.3-2.0 & &  0-285 & 1.25-2.2, 2.8-3.7 \\
10 & 1.25 & 49-51 & 200 & &  95 & -0.775 & 80 & 1.95 & 1.45 & &   0-145 &  1.4-2.0 & &  0-225 &  1.25-2.2, 3.1-3.2 \\
10 & 1.25 & 49-51 & 130 & &  90 & -0.750 & 80 & 1.85 & 1.45 & &   0-125 &  1.4-2.0 & &  0-195 &  1.25-2.1 \\
10 & 1.25 & 49-51 & 60  & &  85 & -0.725 & 80 & 1.85 & 1.45 & &   0-95  &  1.3-1.9 & &  0-155 &  1.25-2.0 \\

30 &  2.1 & 0-100 & $\infty$ & & 130 & -0.925 & 70 & 2.35 & 1.25 & & 105-315 &  2.1-3.2 & & 75-485 &  2.1-4.1 \\
30 &  2.1 & 0-100 & 200 & & 120 & -0.900 & 65 & 2.25 & 1.25 & &  95-235 &  2.1-2.7 & & 75-325 &  2.1-3.3 \\
30 &  2.1 & 0-100 & 130 & & 120 & -0.900 & 65 & 2.25 & 1.25 & &  85-195 &  2.1-2.6 & & 75-275 &  2.1-3.0 \\
30 &  2.1 & 0-100 & 60  & &  80 & -0.950 & 15 & 2.15 & 1.15 & &  75-145 &  2.1-2.3 & & 75-195 &  2.1-2.5 \\

30 &  2.1 & 30-70 & $\infty$ & & 140 & -0.950 & 15 & 2.75 & 1.25 & & 105-285 &  2.1-3.1 & & 75-405 & 2.1-3.9  \\
30 &  2.1 & 30-70 & 200 & & 140 & -0.950 & 15 & 2.75 & 1.25 & &  95-255 &  2.1-2.9 & & 75-355 &  2.1-3.6 \\
30 &  2.1 & 30-70 & 130 & & 120 & -0.875 & 75 & 2.25 & 1.45 & &  85-225 &  2.1-2.8 & & 65-315 &  2.1-3.3 \\
30 &  2.1 & 30-70 & 60  & &  75 & -0.950 & 15 & 2.15 & 1.25 & &  75-165 &  2.1-2.4 & & 65-215 &  2.1-2.7 \\

30 &  2.1 & 49-51 & $\infty$ & & 175 & -0.950 & 15 & 3.15 & 1.25 & &  85-285 &  2.1-3.2 & & 65-405 &  2.1-3.8 \\
30 &  2.1 & 49-51 & 200 & & 120 & -0.925 & 75 & 2.15 & 1.55 & &  85-255 &  2.1-3.1 & & 65-345 &  2.1-3.6 \\
30 &  2.1 & 49-51 & 130 & & 120 & -0.925 & 75 & 2.15 & 1.55 & &  75-225 &  2.1-2.5, 2.7-3.1 & & 65-315 &  2.1-3.3 \\
30 &  2.1 & 49-51 & 60  & &  75 & -0.950 & 40 & 2.15 & 1.25 & &  65-155 &  2.1-2.4 & & 65-225 &  2.1-2.8 \\

30 & 1.25 & 0-100 & $\infty$ & &  80 & -0.025 & 15 & 1.45 & 1.15 & &   0-195 & 1.25-2.3 & &  0-375 &  1.25-3.3 \\
30 & 1.25 & 0-100 & 200 & &  80 & -0.025 & 15 & 1.45 & 1.15 & &   0-135 & 1.25-1.9 & &  0-255 &  1.25-2.6 \\
30 & 1.25 & 0-100 & 130 & &  80 & -0.025 & 15 & 1.45 & 1.15 & &   0-105 &  1.3-1.8 & &  0-195 &  1.25-2.2 \\
30 & 1.25 & 0-100 & 60  & &  70 & -0.025 & 10 & 1.45 & 1.15 & &   0-65  &  1.3-1.6 & &  0-105 &  1.25-1.7 \\

30 & 1.25 & 30-70 & $\infty$ & &  75 & -0.025 & 20 & 1.45 & 1.25 & &   0-175 & 1.25-2.2 & &  0-315 &  1.25-3.1 \\
30 & 1.25 & 30-70 & 200 & &  75 & -0.025 & 20 & 1.45 & 1.25 & &   0-135 & 1.25-1.9 & &  0-265 &  1.25-2.8 \\
30 & 1.25 & 30-70 & 130 & &  75 & -0.025 & 20 & 1.45 & 1.25 & &   0-115 & 1.25-1.8 & &  0-215 &  1.25-2.5 \\
30 & 1.25 & 30-70 & 60  & &  75 & -0.025 & 20 & 1.45 & 1.25 & &   0-75  &  1.3-1.7 & &  0-125 &  1.25-1.9 \\

30 & 1.25 & 49-51 & $\infty$ & &  90 & -0.025 & 10 & 1.45 & 1.25 & &   0-155 & 1.25-2.2 & &  0-305 & 1.25-3.2 \\
30 & 1.25 & 49-51 & 200 & &  90 & -0.025 & 10 & 1.45 & 1.25 & &   0-125 & 1.25-1.9 & &  0-255 & 1.25-2.5, 2.9-2.1 \\
30 & 1.25 & 49-51 & 130 & &  90 & -0.025 & 10 & 1.45 & 1.25 & &   0-105 &  1.3-1.8 & &  0-215 & 1.25-2.4 \\
30 & 1.25 & 49-51 & 60  & &  90 & -0.025 & 10 & 1.45 & 1.25 & &   0-85  &  1.3-1.7 & &  0-125 & 1.25-1.8 \\

\end{tabular}
\end{ruledtabular}
\end{table*}

\begin{figure*}
\resizebox{14.5cm}{!}{\includegraphics{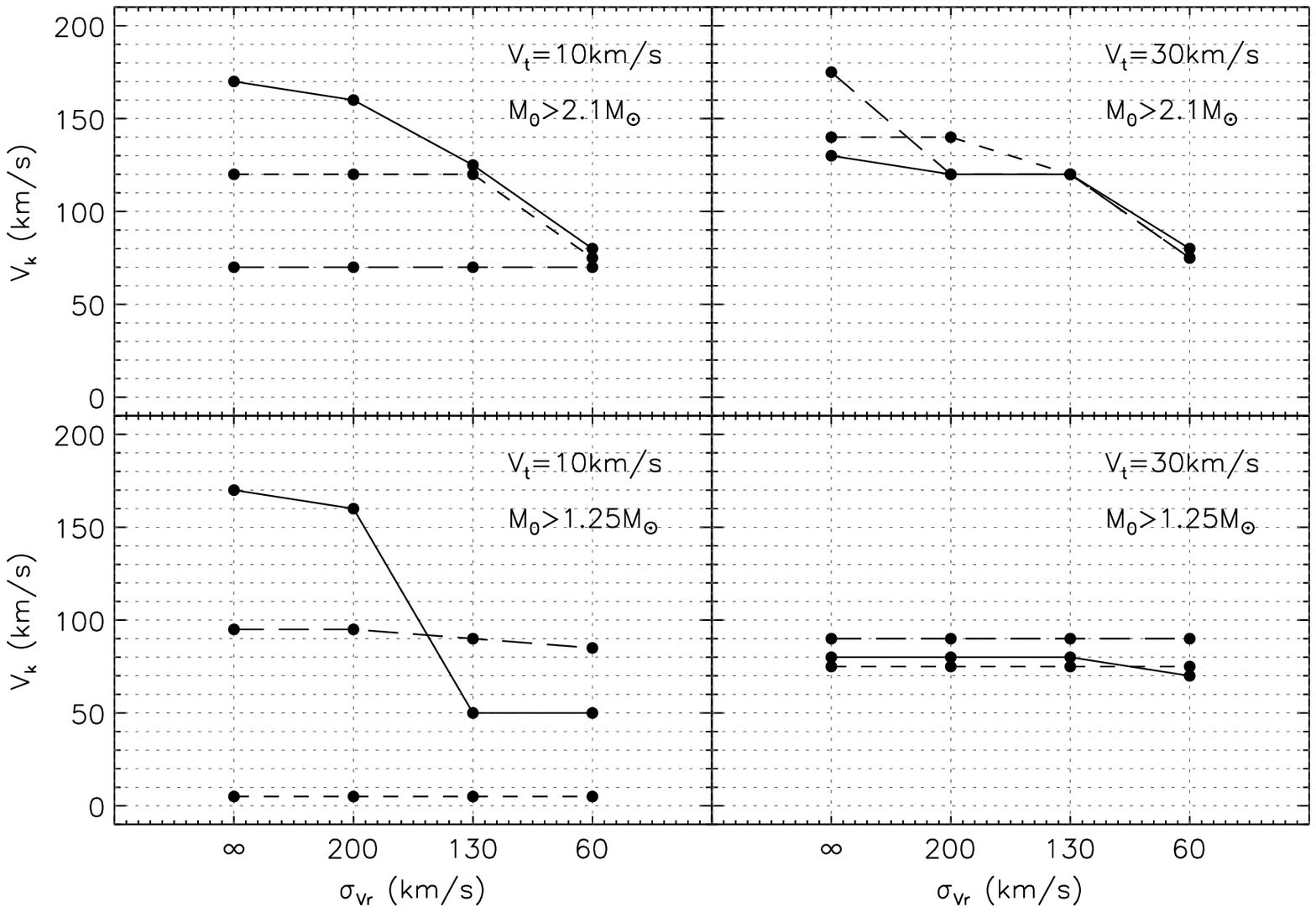}}
\caption{Most likely SN kick velocity magnitude $V_k$ for different
  sets of prior assumptions, for three age ranges: 0--100\,Myr (solid
  lines), 30--70\,Myr (short-dashed lines), 49-51\,Myr (long-dashed
  lines). For brevity, the uniform radial velocity distribution is
  denoted by $\sigma_{V_r}=\infty$.}
\label{modelsVk}
\end{figure*}

\begin{figure*}
\resizebox{14.5cm}{!}{\includegraphics{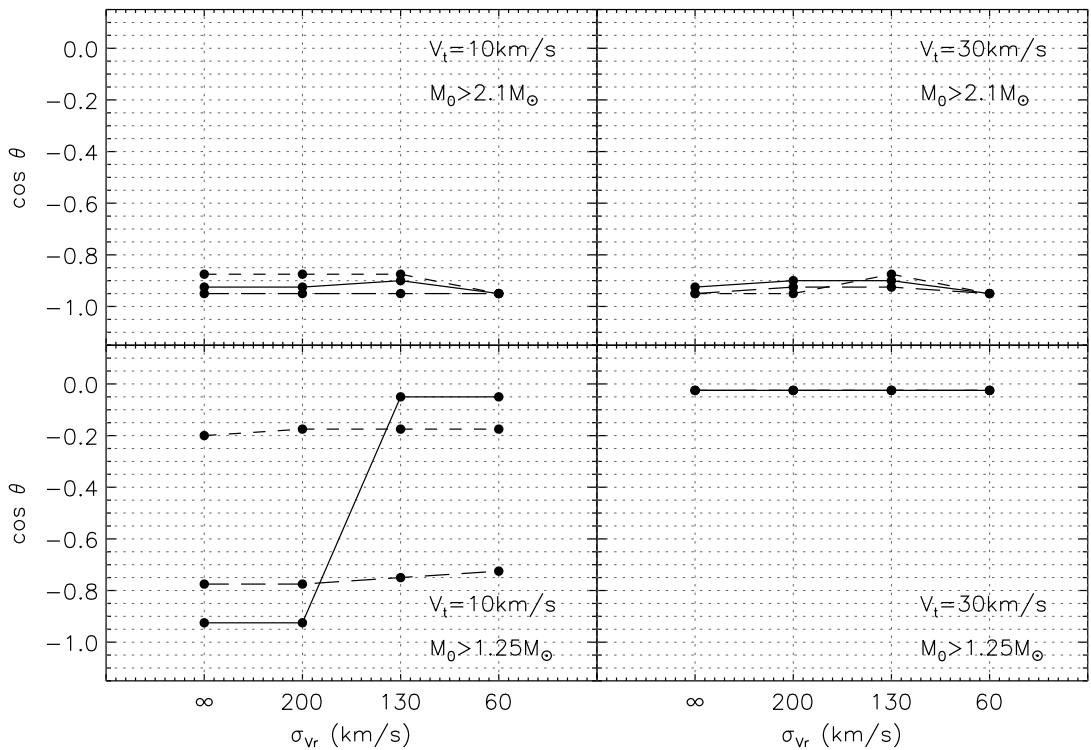}}
\caption{Most likely SN kick direction $\cos \theta$ for different
  sets of prior assumptions, for three age ranges: 0--100\,Myr (solid
  lines), 30--70\,Myr (short-dashed lines), 49-51\,Myr (long-dashed
  lines). For brevity, the uniform radial velocity distribution is
  denoted by $\sigma_{V_r}=\infty$.}
\label{modelsTheta}
\end{figure*}

\begin{figure*}
\resizebox{14.5cm}{!}{\includegraphics{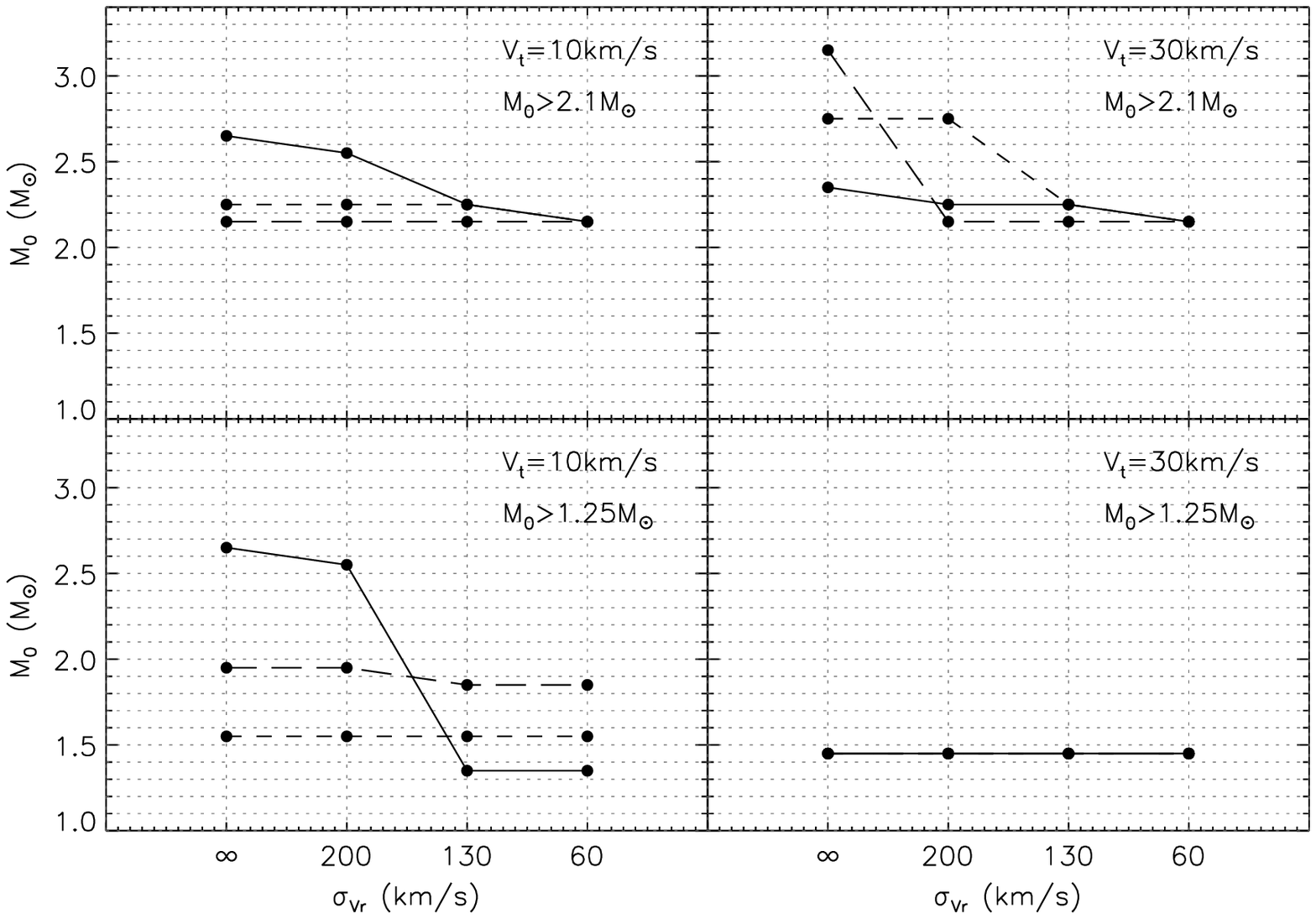}}
\caption{Most likely pre-SN progenitor mass $M_0$ for different sets
  of prior assumptions, for three age ranges: 0--100\,Myr (solid
  lines), 30--70\,Myr (short-dashed lines), 49-51\,Myr (long-dashed
  lines). For brevity, the uniform radial velocity distribution is
  denoted by $\sigma_{V_r}=\infty$.}
\label{modelsM0}
\end{figure*}

The most likely kick velocity and progenitor parameters can also
depend strongly on the assumptions underlying the reconstruction of
the binary's evolutionary history (i.e., the magnitude of the
transverse systemic velocity component, the minimum helium star mass
required for the formation of a NS, the age of the system, and the
probability distribution of the unknown radial velocity). In order to
investigate this, we systematically calculated the 5-D PDFs for $V_k$,
$\cos \theta$, $\phi$, $M_0$, and $A_0$ for different sets of prior
assumptions. The resulting most likely kick velocity and progenitor
parameters are summarized in Table~\ref{pdfmax}. The dependence of the most likely 
$V_k$, $\cos \theta$, and $M_0$ on the adopted prior assumptions is
also presented graphically in
Figs.~\ref{modelsVk}--\ref{modelsM0}. For brevity, we denote the
uniform radial velocity distribution in the table and figures as a
Gaussian distribution with an infinite radial velocity dispersion. 

The most likely kick velocity imparted to pulsar~B at birth is smaller
than 50\,km\,s$^{-1}$ only when the transverse velocity component has
a magnitude of 10\,km\,s$^{-1}$, the minimum helium star mass required
for NS formation is allowed to be as low as $1.25\,M_\odot$, and the
age of the system is assumed to be between 30 and 70\,Myr. {\em All
other models} yield most likely kick velocities of
50--180\,km\,s$^{-1}$. When a minimum pre-SN helium star mass of
$2.1\,M_\odot$ is imposed, the kicks are always strongly favored to be
directed opposite to the helium star's pre-SN orbital motion (most
likely $\cos \theta \simeq -0.90 \pm 0.05$). When the constraint on
the minimum pre-SN helium star mass is relaxed, the most likely kick
direction can shift significantly and can even become perpendicular to
the helium star's pre-SN orbital velocity. Allowing pre-SN helium star
masses down to $1.25\,M_\odot$ furthermore always leads to most likely
progenitor masses of $1.3-2.0\,M_\odot$, {\em except} when
$V_t=10\,{\rm km\,s^{-1}}$, the age of the system is between 0 and
100\,Myr, and the radial velocities are distributed uniformly or
according to a Gaussian with a velocity dispersion of
200\,km\,s$^{-1}$. 

In summary, while small kick velocities of just a few tens of
km\,s$^{-1}$ could be favored for some models, the majority of the
models yields most likely values of 50--180\,km\,s$^{-1}$. Progenitor
masses below $2.1\,M_\odot$ are furthermore not required to explain
the system properties, although they are generally favored when helium
stars below $2.1\,M_\odot$ are still assumed to be viable NS
progenitors.  We note though that the results presented in
Table~\ref{pdfmax} and the associated figures are based on the
assumption that all kick velocity magnitudes $V_k$ are equally
probable. This assumption is inconsistent for models using the
Gaussian radial velocity distributions based on population synthesis
calculations of coalescing DNSs. In particular, these calculations all
adopt a Maxwellian rather than a uniform kick velocity
distribution. In all cases, weighing the kick velocities according to
the Maxwellian underlying the derivation of the radial velocity
distributions would, however, shift the most likely kick velocities to
higher $V_k$ values than listed in Table~\ref{pdfmax}. This reinforces
our conclusion that the presently known observational constraints not
necessarily disfavor kick velocity magnitudes of 100\,km\,s$^{-1}$ or
more.

\subsection{Confidence limits on the kick velocity and progenitor
  parameters}

The use of the full 5-D PDF for $V_k$, $\cos \theta$, $\phi$, $M_0$,
and $A_0$ in the derivation of the most likely pulsar~B kick velocity
and progenitor parameters has the advantage of being free of
projection effects, but is much more cumbersome and computationally
expensive for the calculation of confidence limits. Since our main aim
in this paper is to present a differential analysis showing the
effects of different model assumptions, we adopt a simpler approach
and determine the confidence limits from marginalized 1-D PDFs. We
recognize, however, that in doing so, projection effects may play a
significant role and any information on possible correlations between
the different parameters will be lost.

\begin{figure*}
\resizebox{14.5cm}{!}{\includegraphics{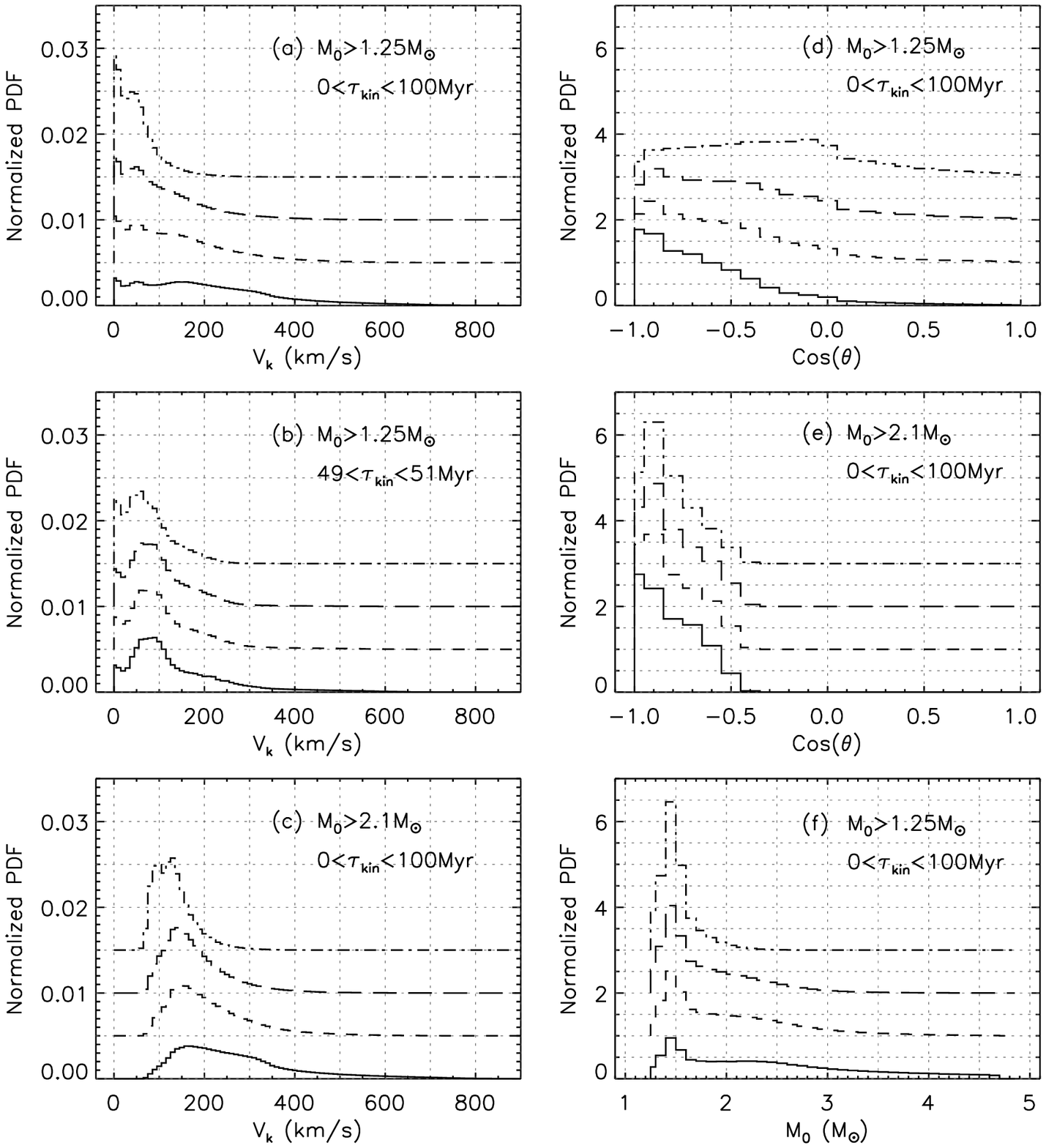}}
\caption{One-dimensional PDFs illustrating some of the dependencies of
  the derived pulsar~B kick velocity and progenitor properties on the
  adopted model assumptions. All plots are for a present-day
  transverse velocity of 10\,km\,s$^{-1}$. Solid lines correspond to
  uniform radial velocity distributions and dashed lines to Gaussian
  distributions with velocity dispersions of 60\,km\,s$^{-1}$
  (long-dashed lines), 130\,km\,s$^{-1}$ (short-dashed line), and
  200\,km\,s$^{-1}$ (dash-dotted line). For clarity, the PDFs are
  offset from each other by an arbitrary amount. Panels~(a)--(c) show
  the distributions of kick velocity magnitudes $V_k$, panels~(d)--(e)
  the distributions of kick direction cosines $\cos \theta$, and
  panel~(f) the distributions of pre-SN helium star masses $M_0$.} 
\label{1DPDF}
\end{figure*}

For illustration, some representative 1-D PDFs used for the
calculation of the confidence limits in the case of a present-day
transverse velocity component of 10\,km\,s$^{-1}$ are shown in
Fig.~\ref{1DPDF}. Panels~(a)--(c) show the kick velocity distributions
resulting from different present-day radial velocity distributions for
ages ranges of 0-100\,Myr and 49-51\,Myr, and minimum pre-SN helium
star masses of $1.25\,M_\odot$ and $2.1\,M\odot$. For a given age
range and minimum pre-SN helium star mass, the PDFs show a peak which
is most pronounced when a Gaussian radial velocity distribution with a
velocity dispersion of 60\,km\,s$^{-1}$ is considered, and which
widens with increasing radial velocity dispersions. When the age is
assumed to be 0-100\,Myr and the minimum pre-SN helium star mass
$1.25\,M_\odot$, there is a clear tendency for the 1-D PDFs to favor
kick velocities of 50\,km\,s$^{-1}$ or less (although this becomes
significantly less pronounced with increasing radial velocity
dispersions). This trend shifts towards favoring kick velocities of
50--100\,km\,s$^{-1}$ when the age range is narrowed to
49-51\,Myr. Moreover, when the age range is kept fixed at 0-100\,Myr,
but the minimum pre-SN helium star mass is increased to
$2.1\,M_\odot$, the favored range of kick velocities shifts to
100--150\,km\,s$^{-1}$. In the latter case, the kick velocity is
furthermore always required to be larger than $\sim 60\,{\rm
km\,s^{-1}}$ (see also Papers~I and~II).

In Panels~(d)--(e), we show the distribution of $\cos \theta$ for
different present-day radial velocity distributions, DNS ages of
0-100\,Myr, and minimum pre-SN helium star masses of $1.25\,M_\odot$
and $2.1\,M_\odot$. For a minimum helium star mass of $1.25\,M_\odot$
and Gaussian radial velocity distributions with velocity dispersions
of more than 130\,km\,s$^{-1}$, the PDFs favor kick directions
opposite to the pre-SN orbital motion. In the case of the Gaussian
radial velocity distribution with a velocity dispersion of
60\,km\,s$^{-1}$, the PDF behaves qualitatively entirely different and
shows a slight preference for kicks perpendicular to the pre-SN
orbital velocity. A non-negligible low-probability tail furthermore
extends well into the region of kick directions in the sense of the
pre-SN orbital motion. When the minimum mass for pulsar~B's pre-SN
helium star progenitor is raised to $2.1\,M_\odot$, the kicks are
always directed opposite to the orbital motion (a lower limit on
$\theta$ of $\sim 113^\circ$ was already derived in Papers~I
and~II). The PDF furthermore becomes fairly insensitive to the adopted
radial velocity distribution.

Panel~(f), finally, shows the distribution of possible pre-SN
progenitor masses for different present-day radial velocity
distributions, DNS ages of 0-100\,Myr, and a minimum pre-SN
helium star mass of $1.25\,M_\odot$. The distributions all favor
progenitor masses of $1.4$--$1.5\,M_\odot$, with the preference for
this mass range being strongest for present-day radial velocity
distributions with small radial velocity dispersions. Distribution
functions for a minimum pre-SN helium star mass of $2.1\,M_\odot$ look
practically the same as the ones displayed panel~(f) if they were cut
off at $2.1\,M_\odot$.

Adopting a present-day transverse velocity of 30\,km\,s$^{-1}$ instead
of 10\,km\,s$^{-1}$ yields very similar conclusions. Overall the
effects of varying model assumptions tend to be somewhat less
pronounced though.

The 68\% and 90\% confidence limits on $V_k$ and $M_0$ obtained from
the 1-D PDFs are listed in Table~\ref{pdfmax} next to the most likely
kick velocity and progenitor parameters
\footnote{Confidence limits for $\cos \theta$, $\phi$, and $A_0$ were
  also calculated but these did not contribute any significant new
  information.}. In the majority of the cases, the confidence limits
contain the most likely $V_k$ and $M_0$ obtained from the full 5-D PDF
for $V_k$, $\cos \theta$, $\phi$, $M_0$, and $A_0$. Whenever this is
not the case, projections effects cause significant shifts between the
values favored by the 1-D and 5-D PDFs. The confidence limits on $V_k$
readily show that the possibility of small kick velocities is entirely
due to the assumption that helium stars less massive than
$2.1\,M_\odot$ can be viable NS progenitors. Under this assumption,
the upper limits on $V_k$ at the 68\% confidence level vary from
65\,km\,s$^{-1}$ to 275\,km\,s$^{-1}$. Thus, even though small kick
velocities become possible, larger kick velocities in excess of
100\,km\,s$^{-1}$ are all but statistically disfavored. This is even
more true when the upper limits at the 90\% confidence level are
considered. For a minimum pre-SN helium star mass of $2.1\,M_\odot$,
the confidence limits for $V_k$ shift to significantly higher values
which are conform with the kick velocity magnitudes derived from
observations of single radio pulsar \citep{2005MNRAS.360..974H,
2005MNRAS.362.1189Z, 2005astro.ph.12585F}. These ``conventional'' kicks are also consistent with the tilt between pulsar~B's rotation axis and the normal to the orbital plane inferred by \citet{2005ApJ...634.1223L}. When the minimum pre-SN
helium star mass is assumed to be $1.25\,M_\odot$, the 68\% and 90\%
confidence limits on $M_0$ typically go all the way down to this
limit, but often also extend upwards to masses larger than
$2.1\,M_\odot$ (especially at the 90\% confidence level). These
numbers confirm our previous conclusion that the presently known
observational constraints not necessarily disfavor progenitors more
massive than $2.1\,M_\odot$ and kick velocities in excess of
100\,km\,s$^{-1}$.

\subsection{Birth sites}

Just like the kick velocity and progenitor parameters, the most likely
birth site of PSR\,J0737-3039B depends strongly on the adopted
transverse systemic velocity component, age range, minimum helium
star mass, and present-day radial velocity distribution. In
\S\,\ref{motion}, we have shown that the distribution of kinematic
ages in the range from 0 to 100\,Myr peaks strongly at ages of
1-2\,Myr (see Fig.~\ref{tauk}). This implies that the system can
travel only a short distance (less than 2\,kpc if the systemic
velocity is less than 1000\,km\,s$^{-1}$) before reaching its current
position. Adopting an age range of 0--100\,Myr therefore yields PDFs
for the birth sites that are strongly peaked near the system's current
position at $X=-0.5$\,kpc and $Y=-8.3$\,kpc, regardless of any of the
other model assumptions.  The situation changes drastically for the
age ranges of 30--70\,Myr and 49--51\,Myr, which exclude very young
ages and therefore impose a minimum distance traveled by the system
since the time it was born.

\begin{figure}
\resizebox{7.5cm}{!}{\includegraphics{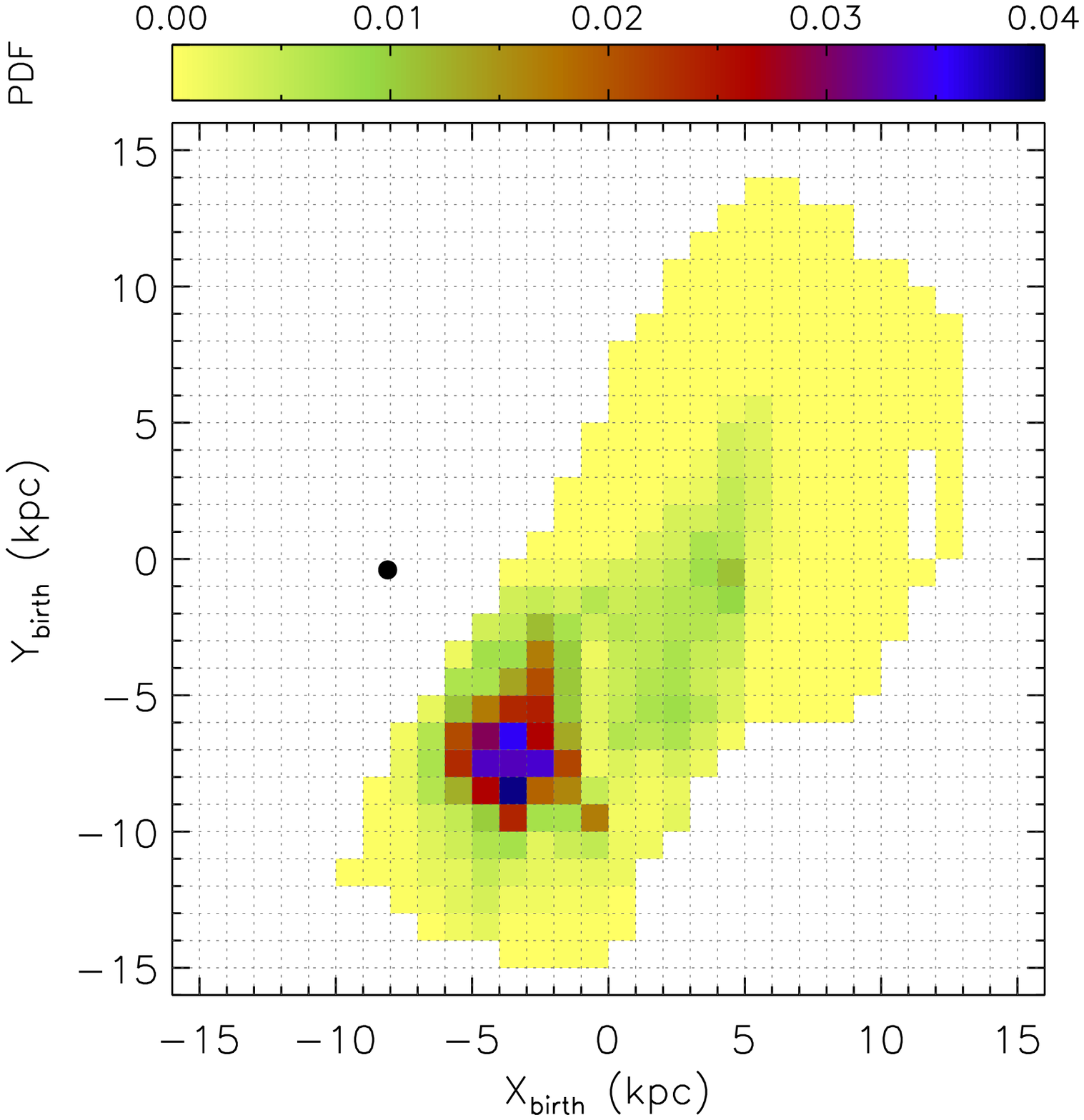}}
\caption{Distribution of possible birth sites of PSR\,J0737-3039 in
  the Galactic plane for a transverse velocity component of $30\,{\rm
  km\,s^{-1}}$, a minimum helium star mass of $1.25\,M_\odot$, an age
  range of 30--70\,Myr, and a Gaussian radial velocity distribution
  with a velocity dispersion of 130\,km\,s$^{-1}$. The solid circle
  indicates the system's present position.}
\label{birth}
\end{figure}

For illustration, the distribution of the possible birth sites in the
Galactic plane is shown in Fig.~\ref{birth} in the case of a
present-day transverse velocity component of $30\,{\rm km\,s^{-1}}$, a
minimum pre-SN helium star mass of $1.25\,M_\odot$, an age range of
30--70\,Myr, and a Gaussian radial velocity distribution with a
velocity dispersion of 130\,km\,s$^{-1}$.  The most likely birth sites
are all located within a radius of 2\,kpc from $\left( X, Y \right) =
\left( -3.5, -7.5 \right)$\,kpc. None of the birth sites are
furthermore close to the system's present-day position indicated by
the solid circle at $\left( X, Y \right) = \left( -0.5, -8.3
\right)$\,kpc. If the minimum helium star mass required for NS
formation is assumed to be $2.1\,M_\odot$ instead of $1.25\,M_\odot$,
the most likely birth sites shown in Fig.~\ref{birth} become entirely
inaccessible. Instead, the binary then most likely originates from a
circle with a radius of 1\,kpc centered on $\left( X, Y \right) =
\left( -2.5, -3.5 \right)$\,kpc, with a clear non-negligible lower
probability tail extending more or less linearly towards $\left( X, Y
\right) = \left( 3.5, 0.5 \right)$\,kpc.

\section{Comparison with previous work}

\subsection{\citet{2004ApJ...616..414W}}

In Paper~II, we derived constraints on the formation and progenitor of
PSR\,J0737-3039B, using the then available observational constraints
on the binary's properties and motion in the Galaxy. The main
differences with the observational and theoretical information adopted
in this investigation are the lower transverse systemic velocity, the
constraint on pulsar~A's spin-orbit misalignment angle, the
possibility of a significantly lower minimum mass for pulsar~B's
pre-SN helium star progenitor, and the use of DNS population synthesis
calculations to obtain theoretical present-day radial-velocity
distributions. Moreover, in Paper~II we mainly focused on the ranges
of possible progenitor parameters, and restricted the statistical
analysis to the derivation of 1-D PDFs for pulsar~B's natal kick
velocity and pulsar~A's spin-orbit misalignment.  Among the
observational and theoretical improvements, the availability of the
constraint on pulsar~A's spin-orbit misalignment has the least effect
on the derivation of the possible pulsar~B progenitor and kick
velocity properties. The main reason for this is that the excluded
range of tilt angles from $60^\circ$ to $120^\circ$ was already
statistically disfavored by the kick velocity and progenitor
constraints derived in Paper~II (see Fig.\ 12 in that paper).

While the majority of the results presented in Paper~II were obtained
assuming a minimum pre-SN helium star mass of $2.1\,M_\odot$, we also
discussed some preliminary test calculations allowing pre-SN helium
star masses below $2.1\,M_\odot$. Because of the strong influence on
the parameter space of the then available proper motion measurement of
$\sim 141\,{\rm km\,s^{-1}}$, we did not find any important changes in
the kick velocity and progenitor constraints related to the decrease
of the minimum admissible helium star mass. With the new proper motion
upper limit of $30\,{\rm km\,s^{-1}}$, the adopted minimum helium star
mass has a much stronger effect on the available parameter space. In
particular, when both the new proper motion constraint and the lower
minimum helium star mass are imposed, there is no longer a lower limit
on the magnitude of the kick velocity imparted to pulsar~B at
birth. Consequently a symmetric SN explosion becomes a viable
formation mechanism for pulsar~B. Moreover, allowing pre-SN helium
star masses smaller than $2.1\,M_\odot$ also removes the lower limit
of $\sim 113^\circ$ on the angle between the kick velocity direction
and the direction of the helium star's pre-SN orbital velocity, so
that the kicks no longer have to be directed opposite to the orbital
motion.

\subsection{\citet{2005PhRvL..94e1102P, 2005astro.ph.10584P}}
\label{PScomp}

More recently, \citet{2005PhRvL..94e1102P, 2005astro.ph.10584P}
(hereafter PS) used a
Monte-Carlo method to argue that the probability of finding the system
close to the Galactic plane is highest if pulsar~B received only a
marginal kick at birth of less than 30\,km\,s$^{-1}$, and if its
pre-SN progenitor had a very low mass of only about
$1.45\,M_\odot$. In their analysis, they considered the post-SN
velocity of the binary randomly projected on the axis vertical to the
Galactic plane (the $Z$-axis introduced in \S\,\ref{motion}) and
followed this one component of the Galactic motion forward in time for
50\,Myr (their assumed age for PSR\,J0737-3039). However, restricting
the Galactic motion to the Z-component can potentially bias the
PDFs, since the horizontal velocity component remains entirely
unconstrained (effectively, they accept all post-SN peculiar
velocities as equally probable regardless of their horizontal
component). In particular, if this horizontal component is large, it
reduces the probability that the system would be found at its current
$X$ and $Y$ position in a similar way as a large vertical component
reduces the probability that the system is found at its current $Z$
position. The analysis of PS furthermore implicitly assumes that the
vertical motion is decoupled from the horizontal one. Although this is
true locally, it is not true globally, since the vertical acceleration
does not depend just on the vertical distance from the plane but also
on the horizontal distance from the Galactic center. Therefore, if the
binary has traveled through a wide range of Galactocentric radii, as
indicated by the large range of possible birth sites displayed in
Fig.~\ref{birth}, its vertical motion is different from the one
modeled by PS.

We investigate this more thoroughly by following the steps outlined by
PS.  Firstly, we determine the post-SN orbital semi-major axis $A$ and
eccentricity $e$ by integrating the equations governing the orbital
evolution due to gravitational wave emission backwards in time, for an
assumed age of 50\,Myr. We find $A=1.42\,R_\odot$ and $e = 0.105$.

Secondly, we randomly generate a pulsar~B kick velocity magnitude
$V_k$, and kick direction angles $\theta$ and $\phi$, assuming a
uniform distribution for the magnitude and an isotropic distribution
for the direction of the kick. Equations~(\ref{eq1}) and (\ref{eq2})
and the associated constraints are then solved for $M_0$ and $A_0$,
using the values of $A$ and $e$ derived by reversing the orbital
evolution for 50\,Myr \footnote{Note that PS consider a range of
eccentricities $0.088 < e < 0.14$ obtained by reversing the past
orbital evolution for a range of ages between 0\,Myr and 210\,Myr (the
characteristic age of {\em pulsar~A}). Although, this is inconsistent
with the age of 50\,Myr adopted for the Galactic motion calculations,
we do not expect the use of the range of eccentricities to have a
strong impact on the derivation of the $V_k$, $\cos \theta$, $\phi$,
$M_0$, and $A_0$ constraints. It is furthermore unclear if PS solve
Eqs.~(\ref{eq1}) and (\ref{eq2}) by randomly generating eccentricities
from a uniform distribution between 0.088 and 0.14, or by some other
unspecified procedure. For these reasons, we modify this step in PS's
analysis and consistently use the post-SN orbital parameters
associated with a fixed age of 50\,Myr.}. If a solution for $M_0$ and
$A_0$ exists, the magnitude of the post-SN peculiar velocity imparted
to the binary's center of mass is calculated by means of
Eq.~(\ref{vpec}).

Thirdly, we approximate the motion of the system in the Galaxy by a
local vertical oscillation with an amplitude drawn from a Gaussian
distribution with a dispersion of 50\,pc. At the time of the SN
explosion, the system is assumed to be at a uniformly distributed
random phase of the oscillation. The vertical motion of the system in
the Galaxy is then followed forward in time for 50\,Myr using the
Galactic potential of \citet{1990ApJ...348..485P}. For this purpose,
the randomly generated starting point has to be supplemented with the
$Z$-component of the calculated post-SN peculiar velocity. Since the
kick direction angles $\theta$ and $\phi$ are defined with respect to
the pre-SN orbital plane, which has an unknown direction in space, the
$Z$-component of the post-SN peculiar velocity is determined by
assigning an isotropically distributed random direction to the
calculated post-SN peculiar velocity. This new direction is generated with respect to the Galactic frame of reference $OXYZ$ introduced in \S\,\ref{motion}, and therefore straightforwardly leads to the determination of the $Z$-component of the post-SN peculiar velocity.

\begin{figure*}
\resizebox{14.5cm}{!}{\includegraphics{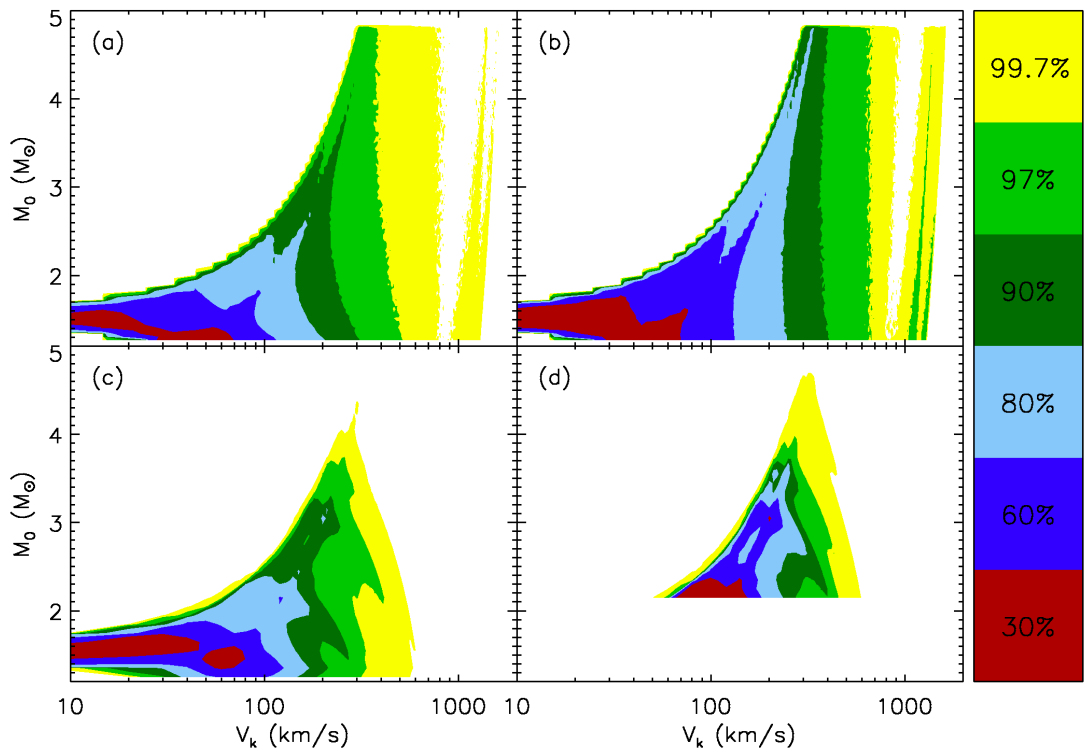}}
\caption{Contour plots for the $\left( V_k, M_0
  \right)$-PDF. Panels~(a) and~(b) correspond to the PDFs obtained
  using the method outlined by PS. The PDF shown in Panel~(a) only
  includes systems with $|Z| \lesssim 50$\,pc, while the PDF shown in
  panel~(b) does not incorporate any constraints on the position of the
  system in the Galaxy. Panels~(c) and~(d) correspond to the PDFs
  obtained using the method outlined in this paper. The PDF shown in
  panel~(c) corresponds to a present-day transverse velocity of
  30\,km\,s$^{-1}$, a minimum pre-SN helium star mass of $1.25\,M_\odot$,
  kinematic ages of 49--51\,Myr, and a Gaussian radial velocity
  distribution with a dispersion of 130\,km\,s$^{-1}$. The PDF shown
  in panel~(d) adopts the same assumptions, except that the minimum
  pre-SN helium star mass is assumed to be $2.1\,M_\odot$.}
\label{ps}
\end{figure*}

If the generated system satisfies all SN orbital dynamics constraints
and if, after following the motion in the Galaxy for 50\,Myr, the
system ends up within 50\,pc of the Galactic mid-plane, it is
considered a possible progenitor of PSR\,J0737-3039
\footnote{Contrary to PS, we do not impose any constraint on the
present-day transverse systemic velocity, since there is no
unambiguous way to impose the constraint without making additional
assumptions about the system's present-day 3-D space velocity. In
particular, limiting the motion of the system in the Galaxy to a local
vertical oscillation restricts the knowledge of the velocity in the
simulations to the component perpendicular to the Galactic
plane. Since this component does not lay in the plane perpendicular to
the line-of-sight to the double pulsar, imposing the proper motion
constraint requires additional assumptions on the radial velocity and
the direction of the proper motion. }. The procedure is then repeated
until $10^8$ such systems are found. The outcome of the simulation is
a 2-D PDF of $M_0$ and $V_k$ shown in panel~(a) of Fig.~\ref{ps}. The
PDF is normalized so that its integral over the entire admissible
parameter space is equal to one. This is different from PS who
normalize their PDF to the peak probability density \footnote{Because
of the chosen normalization, the contour levels in PS measure the
decrease of the PDF with respect to its maximum value. They therefore
do not correspond to confidence levels.}. The overall features of this
PDF are to be compared to the upper right panel of Fig.\ 1 in
\citet{2005PhRvL..94e1102P}; it is evident that the two PDFs are in
rather good agreement.  Differences in the details can be partly
attributed to the PDF normalization adopted by PS (since it requires that the {\em exact} PDF peak is ``captured'' by the Monte Carlo process), which
depends on the accuracy of the Monte-Carlo simulation and the size of
the bins used to construct the PDF.

In order to assess the effect of approximating the space motion of
PSR\,J0737-3039 as a vertical oscillation centered on the system's
current position, we repeated the above calculation for a system
completely identical to the double pulsar, but located at a different
position in the Galactic plane. In particular, we positioned the
system closer to the Galactic center at $X=-2.5$\,kpc and
$Y=-3.5$\,kpc (this corresponds to the most likely birth site found in
\S\,\ref{results} for a present-day transverse velocity of
30\,km\,s$^{-1}$, an age of 30--70\,Myr, a Gaussian radial velocity
distribution with a velocity dispersion of 130\,km\,s$^{-1}$, and a
minimum pre-SN helium star mass of $2.1\,M_\odot$). Somewhat
surprisingly, we found the effect on the PDFs to be rather minimal. As
a second test, we therefore considered a second system identical to
the double pulsar, but located at a height of $790 \pm 50$\,pc above
the Galactic plane (corresponding to the distance of PSR\,B1534+12 to
the Galactic plane). Again, we found minimal changes in the PDF. We
conclude that the position of the system in the Galaxy therefore seems
to be relatively unimportant in the derivation of the $\left( V_k, M_0
\right)$-PDF.

This conclusion is further strengthened by the results in panel~(b) of
Fig.~\ref{ps}: it shows the 2-D PDF obtained {\em without any}
constraint on $|Z|$. It is evident that the PDF does not depend in any
significant way on the position of PSR\,J0737-3039 close to the
Galactic plane, in contrast to what is claimed by PS (the main
difference between panels~(a) and~(b) is that the "peak" contour is
somewhat more spread out). Indeed when tracking the progenitor systems
as they are affected by the different constraints in the simulations,
we find that the $|Z| < 50$\,pc constraint eliminates only 13\% of the
systems that satisfy all other imposed constraints. It turns out that
the orbital dynamics of asymmetric SN explosions and associated
constraints are much more important than the position of the system in
the Galaxy in determining the $\left( V_k, M_0 \right)$-PDF.

For comparison, we also constructed 2-D kick velocity and progenitor
mass distributions by marginalizing the 5-D PDFs derived in
\S\,\ref{results}, for model assumptions similar to those adopted by
\citet{2005astro.ph.10584P}. Specifically, we considered a present-day
transverse velocity of 30\,km\,s$^{-1}$, a minimum helium star mass of
$1.25\,M_\odot$, kinematic ages of 49--51\,Myr, and a Gaussian radial
velocity distribution with a dispersion of 130\,km\,s$^{-1}$. We also
removed the limits on pulsar~A's spin-orbit misalignment from the list
of imposed constraints as it was not incorporated in the analysis of
PS. The resulting PDF is shown in panel~(c) of Fig.~\ref{ps}. We note
that, although the analysis by PS does not incorporate the radial
velocity, it is an indispensable part of our method which cannot be
omitted from the analysis. A comparison between our results and those
derived by PS with exactly the same assumptions is therefore not
possible. Despite this, the differences between the PDFs displayed in
panels~(a) and~(c) are quite minimal. The same holds true if the Gaussian radial
velocity distribution with a dispersion of 130\,km\,s$^{-1}$ is replaced by any of the
other radial-velocity distributions considered. In view of the
non-negligible differences between the two approaches, this agreement
is all but trivial and thus strengthens our confidence in the validity
of the derived results. The effect of the minimum helium star mass on
the PDF is illustrated in panel~(d), where the minimum pre-SN
helium star mass was assumed to be $2.1\,M_\odot$. The main difference 
with the PDF shown in panel~(c) is the minimum kick velocity of $\sim
60$\,km\,s$^{-1}$. The most likely kick velocity is of the order of
60--100\,km\,s$^{-1}$ and the most likely progenitor mass is
$2.1-2.4\,M_\odot$.

Thus, we conclude that (i) if one allows {\em a priori} for low-mass
progenitors of pulsar B, and (ii) if the marginalized 2-D PDF is
considered and not the underlying 5-D PDF, then the preference for low
progenitor masses and low kicks is rather generic and it is not
related to the system's position close to the plane or the small upper
limit on the proper motion in any crucial way (c.f. PS). This applies to both the analysis presented in this paper and the analysis by PS. Examining the role of the prior assumptions and considering the full 5-D PDF of $V_k$, $\cos \theta$, $\phi$, $M_0$, and $A_0$ are therefore essential in establishing an unbiased view of the possible formation mechanism and progenitor of PSR\,J0737-3939B.

Finally, we note that, contrary to what is claimed by PS, small tilt angles between pulsar
A's spin angular momentum and the post-SN orbital angular momentum do
not necessarily imply that a small SN kick was imparted to pulsar~B at
birth. This can, e.g., be seen from Figs. 11 and 12 in Paper~II. In
addition, Eq.~(\ref{tilt}) shows that the tilt depends on the
direction as well as on the magnitude of the natal kick velocity, and
that in the particular case where the kicks are confined to the pre-SN
orbital plane, the tilt angle is equal to zero regardless of the
magnitude of the kick velocity.

\section{Summary and concluding remarks}

In this study, we have considered the kinematic and evolutionary
history of the double pulsar PSR\,J0737-3039, taking into account the
most up-to-date observational constraints. Unlike our earlier work on this topic, we focus here on
the derivation of the full multi-dimensional probability distribution
functions for the magnitude and direction of the kick velocity
imparted to pulsar~B at birth, the mass of pulsar~B's pre-SN helium
star progenitor, and the pre-SN orbital separation. The consideration
of the multi-dimensional PDF has the distinct advantage that it is
free of projection effects and that it accounts for all possible
correlations between the derived parameters; these correlations
and projection effects turn out to be very important when addressing
the question of what are the most likely properties of the double
pulsar's progenitor. We also examine the dependence of the PDFs on
the prior assumptions for the magnitude
of the transverse systemic velocity component, the minimum pre-SN helium star mass required to form a NS, the age of the system, and the present-day radial velocity. Since the latter cannot be determined from observation, we construct and explore theoretical radial-velocity distributions by means of population synthesis calculations for coalescing DNSs.

One of the main results of our analysis is that marginalizing the full five-dimensional progenitor PDF to 1-D or 2-D distributions for the pulsar~B kick velocity and progenitor mass has a major effect on the determination of their most likely values. When the full multi-dimensional PDF is examined, it is clear that although some sets of prior assumptions indeed favor low kick velocities and low progenitor masses, as claimed by \citet{2005PhRvL..94e1102P}, the majority of the models favor kick velocities of 50--180\,km\,s$^{-1}$ and progenitor masses of 1.45--2.75\,$M_\odot$. 

In particular, if the transverse systemic velocity is assumed to be 30\,km/s and helium stars less massive than 2.1\,$M_\odot$ are assumed to be viable NS progenitors, the most likely pulsar B progenitor mass is 1.45\,$M_\odot$ regardless of any of the other prior assumptions. If, on the other hand, the transverse systemic velocity is assumed to be 10\,km/s while  keeping the assumption that helium stars less massive than 2.1\,$M_\odot$ are viable NS progenitors, the most likely progenitor mass can vary from  1.35\,$M_\odot$ to 2.65\,$M_\odot$, depending on the assumed systemic age and radial velocity distribution. Hence, whether progenitor masses greater than 2.1\,$M_\odot$ are statistically likely or unlikely depends strongly on the adopted prior assumptions (cf. \citep{2005PhRvL..94e1102P, 2005astro.ph.10584P}). Most likely progenitors with $M_0 < 2.1\,M_\odot$ can furthermore also be associated with kick velocities of up to 100\,km\,s$^{-1}$ (see, e.g., the case of $V_t=30\,{\rm km\,s^{-1}}$ and $M_0 > 1.25\,M_\odot$ in Figs.~\ref{modelsVk} and~\ref{modelsM0}), while kick velocities of less than $60\,{\rm km\,s^{-1}}$ are only possible for progenitor masses below $2.1\,M_\odot$.

We also find that the proximity of PSR\,J0737-3039 to the
Galactic plane and the small proper motion do not pose stringent
constraints on the kick velocity and progenitor mass of
pulsar~B. Instead, the constraints are predominantly determined by the
orbital dynamics of asymmetric SN explosions.  This is in contrast to the work of \citet{2005PhRvL..94e1102P, 2005astro.ph.10584P} who emphasize that the proximity of the double pulsar to the Galactic plane implies that pulsar~B most likely received only a small kick at birth and that its progenitor most likely had mass of  $\sim 1.45\,M_\odot$.

Hence, based on the currently available observational constraints, a
wide range of progenitor and kick velocity properties are favored for
PSR\,J0737-3039B. In particular, the constraints are compatible with a
conventional hydrodynamical or neutrino-driven SN explosion from a
helium star more massive than $2.1\,M_\odot$
\citep{1986A&A...167...61H, 2003astro.ph..3456T}, as well as an
electron-capture SN from a helium star less massive than
$2.5\,M_\odot$ \citep{1984ApJ...277..791N,
1987ApJ...322..206N}. \citet{2005MNRAS.361.1243P} have speculated that
the electron-capture SN mechanism may be typical for close binaries
and that it may be accompanied by much smaller kicks than
hydrodynamical or neutrino-driven SN explosions. Consequently, if
pulsar~B is formed through an electron capture SN and if electron
capture SNe are accompanied by small kicks, the mass of pulsar~B's
pre-SN helium star progenitor must be smaller than $2.1\,M_\odot$
(otherwise the kick is always larger than $\sim 60\,{\rm
km\,s^{-1}}$). Since it is unlikely that future observations will lead to new constraints on the smallest possible pulsar~B progenitor mass,
further insight to the formation mechanism of pulsar~B should be
sought in core-collapse simulations of low-mass ($\lesssim
2.1\,M_\odot$) helium stars and population synthesis studies of
PSR\,J0737-3039-type binaries and their progenitors.

\acknowledgments We are grateful to Laura Blecha for sharing the code
used to calculate the Galactic motion of PSR\,J0737-3039 backwards in
time, and to Richard O' Shaughnessy for useful discussions on
multi-dimensional statistics.  This work is partially supported by a
NSF Gravitational Physics grant PHY-0353111, a David and Lucile
Packard Foundation Fellowship in Science and Engineering grant, and
NASA ATP grant NAG5-13236 to VK; and a Northwestern University Summer
Research Grant to JK. KB acknowledges the support of KBN grant 1P03D02228.

\appendix

\section{Comments by Piran \& Shaviv on this work}

\citet{2006astro.ph..3649P} recently posted a Comment on the astro-ph preprint server "refuting" the work presented in this paper. Their argumentation, however, rests on severe misstatements of our analysis and neglects some of the major conclusions drawn in the previous sections. While we have already addressed their criticism at appropriate places throughout this paper, we here wish to summarize our counterarguments to eliminate any further misunderstandings:

\medskip

(1) Our results are not "in spite of the observations" as claimed by \citet{2006astro.ph..3649P}. They are based on all currently available observational constraints, most of which are in fact initial conditions in our calculation. 

\medskip

(2) While we do conclude that most of our models yield kick velocities of 50--180\,km\,s$^{-1}$, we do not claim that the models favoring lower kick velocities are less likely. We conclude even less that the most likely pulsar B progenitor mass is large. It can, e.g., be seen from Table~\ref{pdfmax} that if the transverse systemic velocity is assumed to be 30\,km/s and the minimum pulsar B progenitor mass is assumed to be 1.25\,$M_\odot$, the most likely progenitor mass is 1.45\,$M_\odot$ regardless of any of the other prior assumptions. If, on the other hand, the transverse systemic velocity is assumed to be 10\,km/s while the minimum progenitor mass is kept at 1.25\,$M_\odot$, the most likely progenitor mass can vary from  1.35\,$M_\odot$ to 2.65\,$M_\odot$, depending on the assumed age range and radial velocity distribution. Our main conclusion is thus that the prior assumptions play an important role in determining the most likely kick velocity and progenitor properties.

\medskip

(3) We do not assume that the system is moving with a very large velocity almost exactly towards. Instead, we adopt theoretical radial-velocity distributions based on realistic binary evolution models for coalescing DNSs that reside close to the Galactic plane and have small transverse systemic velocity components, consistent with the current observational constraints on the kinematics of the double pulsar. Because of the latter constraints, the distributions implictly incorporate the geometrical probability argument put forward in Eq. (2) of \citep{2006astro.ph..3649P}. We furthermore also consider four different distribution functions (one flat distribution and three Gaussians of varying width) in order to assess the robustness of our results (see, e.g., Table~\ref{pdfmax}). 

\medskip

(4) The differences between our conclusions and those of \citet{2005PhRvL..94e1102P} rest on the consideration of the full 5-D PDF instead of a marginalized 2-D one, not on the modeling of the motion of the system in Galaxy. The reason for this is that the orbital parameters and the dynamics of asymmetric SN explosions turn out to be much more constraining than the system's position and motion in the Galaxy (contrary to what is claimed by \citet{2005PhRvL..94e1102P}).
\citet{2006astro.ph..3649P} also completely miss that when we adopt similar assumptions as they do, and marginalize our 5-D PDFs for $V_k$, $\cos \theta$, $\phi$, $M_0$, and $A_0$ to 2-D PDFs for $M_0$ and $V_k$, we find results that are in excellent agreement with theirs, irrespective of the treatment of the unknown radial velocity. 

\medskip

(5) The population synthesis calculations used to construct the theoretical radial-velocity distributions incorporate the possibility that mass transfer takes place right before the formation of pulsar~B. The models therefore allow the pre-SN masses of helium stars in DNS progenitors to be as low as the $1.45\,M_\odot$ pulsar~B progenitor mass favored by \citet{2005PhRvL..94e1102P}. The derived radial-velocity distributions are thus by no means biased towards high progenitor masses as claimed by \citet{2006astro.ph..3649P}. 

\medskip

(6) Based on the preceding arguments, the conclusion of \citet{2006astro.ph..3649P} that "we assume the result we obtain" clearly rests on severe misinterpretations of the calculations underlying our analysis. 

\medskip

We want to conclude this discussion by stressing that the goal of our paper is not to refute the progenitor and kick constraints derived by \citet{2005PhRvL..94e1102P}. We do however want to make it clear that there are many additional factors that were not considered by these authors, and that these may affect the determination of the most likely progenitor and kick velocity properties.

\end{document}